%% file: main.tex
\documentclass[10pt,conference]{IEEEtran}
\usepackage{cite}
\usepackage{amsmath,amssymb,amsfonts}
\usepackage{algorithm}
\usepackage{algorithmic}
\usepackage{graphicx}
\usepackage{textcomp}
\usepackage{xcolor}
\usepackage[hyphens]{url}
\usepackage{fancyhdr}
\usepackage{hyperref}

\usepackage{multirow}
\usepackage{array}
\newcolumntype{P}[1]{>{\centering\arraybackslash}p{#1}}
\usepackage{makecell}
\usepackage{mathtools}
\usepackage{braket}
\usepackage{pifont}
\usepackage[flushleft]{threeparttable}
\usepackage{enumitem}
\usepackage{tikz}
\usepackage{amsthm}

\title{\methodName: Scheduling of Distillation and Coding for
\underline{R}ate-\underline{Fi}delity Tradeoff in Quantum \underline{Ne}tworks}

\long\def\ignore#1{}
\def \methodName{ReFINE}
\def \approachOne{\mbox{\methodName{}-M}}
\def \approachZero{\mbox{\methodName{}-D}}
\def \approachTwo{\mbox{\methodName{}-C}}
\def\LogicalQ{CEM}

\newcommand{\new}[1]{#1}

\author{
  Narges Alavisamani$^*$,
  Matthieu Bloch, 
  Moinuddin Qureshi
  \\[0.2cm] 
  Georgia Institute of Technology
}

\ifdefined\eaopen

\fi
\begin{document}
\maketitle
\let\thefootnote\relax\footnotetext{\hspace{-11pt}$^*$The corresponding author can be reached at narges.alavisamani@gatech.edu}


\input{./sections/0_abstract}

\input{./sections/1_introduction}

\input{./sections/2_background}

\input{./sections/3_evaluation_methodology}

\input{./sections/4_model}
\input{./sections/5_evaluations}

\input{./sections/6_related_work}
\input{./sections/8_conclusion}
\input{./sections/acknowledgement}
\input{./sections/9_appendix}


\bibliographystyle{IEEEtranS}
\bibliography{refs}

\end{document}

%% file: sections/0_abstract.tex
\begin{abstract}

In quantum networks, nodes are connected via sharing of Einstein– Podolsky-Rosen (EPR) pairs, ideally with high {\em fidelity} and high {\em rate}. 
However, the fidelity of EPR pairs degrades due to imperfect generation and decoherence errors, requiring compensating methods to maintain acceptable levels of fidelity. 
{\em Entanglement Distillation} is a method that increases the fidelity but operates probabilistically and may destroy all involved EPR pairs upon failure. This failure reduces available EPR pairs for application use, thereby decreasing the service rate. Quantum Error Correction (QEC) is another mechanism to protect EPR pairs against error by forming what we term as {\em Coding-Enhanced Memory (CEM)}. While effective, CEM requires extra time and resources to form the code, which also reduces the service rate. Existing methods often use static combinations of distillation and CEM, ignoring demand variations. This results in a low service rate without significant fidelity gain. Limited resources together with this rate-fidelity tradeoff make it essential to {\em schedule} when to run distillation, form CEM, or serve requests.

We propose {\em ReFINE}, a {\em demand-aware preemptive scheduler} that based on application requirements either serves an available EPR pair immediately or preserves it in CEM. This selective use of CEM, only when needed, enables a better balance for rate-fidelity tradeoff than always using CEM. Between request arrivals, ReFINE either schedules distilling EPR pairs or forming CEM to protect distilled pairs, following one of the three priority policies: {\em ReFINE-D (Distillation-First)} first generates EPR pairs for distillation and then forms the CEM, prioritizing service rate. {\em ReFINE-M (Memory-First)} first forms the CEM, then generates the EPR pairs for distillation, prioritizing fidelity. {\em ReFINE-C (Concurrent)} performs both distillation and CEM formation concurrently, balancing between fidelity and service rate. Across diverse demand patterns, our evaluations show that ReFINE closes up to 11.21\% of the gap to the ideal (unit) fidelity compared to continuous distillation, and up to 49.48\% of the gap to ideal service rate compared to the static methods.

\end{abstract}

%% file: sections/1_introduction.tex
\section{Introduction}

Quantum networks use entanglement to enable applications such as key distribution, sensing, and distributed computing~\cite{wu2022autocomm, wu2023qucomm, zhang2024mech, de2024thresholds, kim2024fault, degen2017quantum}. Fundamental to the operation of a quantum network is distribution of entanglement, such as {\em Einstein–Podolsky–Rosen (EPR)} pairs~\cite{Nielsen_Chuang_2010}, between two nodes (``Sender" and ``Receiver" nodes)~\cite{wehner2018quantum}. 


\noindent{\bf{Rate-Fidelity Tradeoff in Quantum Networks:}} To share an EPR pair between two nodes, the sender node generates the EPR pair and sends one of the qubits of the EPR pair to a receiver node. In our model, similar to prior work~\cite{chakraborty2019distributed, dupuy2023survey}, the nodes proactively share EPR pairs (before EPR pairs are requested to be consumed for operations). If a node receives a request but has no EPR pair ready, it declines the request.
Ideally, we want the nodes in a quantum network to share EPR pairs of high {\em fidelity} at high {\em rate} (equivalently, low probability of request getting declined). Fidelity~\cite{Nielsen_Chuang_2010, wehner2018quantum} measures the closeness of the noisy EPR pair state to the noiseless EPR pair state.  We term the rate at which a node serves incoming requests for EPR pairs as the node {\em service rate}.

\begin{figure*}[th]
	\centering
	\includegraphics[width=0.93\textwidth]{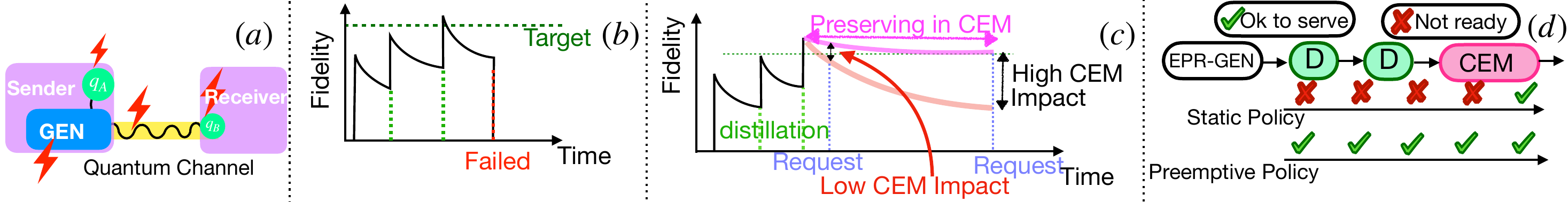}
    \caption{(a) Entanglement distribution: Sender generates an EPR pair, sends one qubit of the pair to Receiver. Errors reduce EPR pair's fidelity. (b) As decoherence reduces fidelity, CED attempts to restore it, but frequent distillation rounds increase failure rates. (c)~Depending on when consumption requests arrive, the utility of CEM varies. (d) \methodName{} schedules limited distillation rounds, preservation in CEM (only if needed), and serves requests preemptively.}
     
	\label{fig:intro_fig}
\end{figure*}

Quantum network applications have different requirements for fidelity and service rate. Some applications require a critical high fidelity, such as BBM92 {\em Quantum Key Distribution}~\cite{bennett2014quantum}, which must maintain fidelity above 0.83 to prevent faked attacks~\cite{sajeed2020bright, vardoyan2023quantum, gottesman2004security, scarani2009security}. Others, such as quantum position verification and long-baseline interferometric telescopes, depend on maintaining a critical minimum service rate~\cite{beigi2011simplified, asadi2025linear, cowperthwaite2023towards, tomamichel2013monogamy}. If the rate drops below this threshold, the application fails regardless of fidelity. In practice, both fidelity and service rate fall short of these critical requirements. Imperfect EPR generation, depolarizing errors, and decoherence over time degrade fidelity (Figure~\ref{fig:intro_fig}(a)). Attenuation and rates of EPR pair generation limit service rate. Our work focuses on equipping nodes in the network with techniques to increase fidelity and rate to match the demand. The existing methods typically boost fidelity at the cost of consuming extra time and resources, which can further reduce the service rate.

\noindent{\bf{Entanglement Distillation:}} This method relies on consuming one or more {\em auxiliary} EPR pairs to improve the fidelity of a {\em target} EPR pair~\cite{deutsch1996quantum, bennett1996purification, kwiat2001experimental, dur1999quantum, yan2022entanglement, yan2023advances, pan2003experimental, hage2008preparation, dong2008experimental, takahashi2010entanglement, hage2010iterative}. In a basic 2-to-1 entanglement distillation protocol, the sender and receiver operate on their halves of the target and auxiliary pairs shared between them, perform local gates (e.g. CNOT), and measure the auxiliary pair.  
They compare the measurement results using a classical channel. Whether the measurements match or not is probabilistic. 
If they match, the distillation succeeds and the fidelity of the target EPR pair is improved.  If not, distillation fails and both target and auxiliary pairs are discarded. Distillation consumes capacity from the EPR pair generator, the core {\em resource} in quantum networks. This reduces the capacity of the generator available to serve requests, lowering the service rate. Moreover, if distillation fails, all invested resources are discarded and the service rate drops further without any fidelity improvement.

\noindent{\bf{Coding-Enhanced Memory:}} We can use QEC codes to maintain the fidelity of EPR pairs for nodes in a quantum network~\cite{jiang2009quantum, munro2010quantum, fowler2010surface,muralidharan2014ultrafast}. We term the fault-tolerant qubit as {\em Coding-Enhanced Memory (\LogicalQ{})}, as it protects half of an EPR pair against errors for a longer duration~\cite{calderbank1996good, dennis2001surfacecodes,fowler2012surface, kitaev1997toriccodes, landahl2011colorcodes, shor1995scheme, steane1996multiple}.  \LogicalQ{} requires additional entangled qubits, which may consume capacity from either the limited {\em resource} of the EPR pair generator shared with link-level EPR generation or from a generator dedicated to CEM. Formation of \LogicalQ{} depends on the specific code.  For example, we consider teleportation-based Shor’s code~\cite{shor1995scheme,luo2021quantum},  which requires GHZ$_4$ state formation (consuming two EPR pairs) to preserve half of a shared EPR pair~\cite{shor1995scheme,dennis2001surfacecodes,luo2021quantum}, though this formation process consumes time that reduces service rate.

\noindent{\bf{Limitations of Existing Policies:}} If a distilled EPR pair is not used immediately, decoherence degrades its fidelity again.
To mitigate this, a common policy,  {\em Continuous Entanglement Distillation (CED)}~\cite{davies2024entanglement, elsayed2023fidelity, mundarain2009entanglement}, distills repeatedly until the EPR pair is consumed (Figure~\ref{fig:intro_fig}(b)). However, the success of each distillation round
is probabilistic, and the likelihood of all rounds
succeed is low. For example, performing three rounds
of distillation on an EPR pair with an initial fidelity of 0.8 fails 54\% of the time ~\cite{davies2024entanglement, bennett1996purification}. Such failure reduces availability
of EPR pairs and service rate.  To avoid failure while preventing decoherence, {\em Static CEM} policy exclusively uses CEM~\cite{muralidharan2016optimal,jiang2009quantum,fowler2010surface,muralidharan2014ultrafast,munro2012quantum} . But, static CEM reduces service rate, due to rejecting interim requests during CEM formation.

While Static CEM only preserves fidelity, some policies combine distillation with CEM in a static manner ({\em Static Distill+CEM}), first increasing fidelity and then preserving it~\cite{patil2024entanglement,muralidharan2016optimal,pathumsoot2024boosting}. In Static Distill+CEM,  a sender node has an EPR pair available only after it has completed the specified number of distillation and has been preserved in the CEM. Any request arriving in the interim (between the generation of EPR pair and preservation in CEM) is declined due to unavailability of the EPR pair (Figure~\ref{fig:intro_fig}(d)). This unavailability reduces service rates, without significantly improving fidelity. All existing policies fix the order of distillation and CEM, and no prior work schedules these operations at the node level. {\em Our goal is to achieve a better balance in the rate–fidelity tradeoff through scheduling and resource allocation.}

\noindent{\bf{Insight:}} The benefit of CEM depends on the duration between successful distillation and request arrival from the receiver (Figure~\ref{fig:intro_fig}(c)). If the duration is small, then the reduction in fidelity due to decoherence is also small, and utility of CEM is low.  On the other hand, if the duration is long, the qubits are more vulnerable to decoherence, and the utility of CEM is high (and we have enough time to create the CEM and preserve the qubit without impacting the service rate). Thus, the decision of whether using
CEM or not should be done dynamically depending on the request arrival timing. 

Resource configuration also interacts with timing in non-intuitive ways, so \textit{adding more resources does not necessarily yield better performance.} For example, in Static Distill+CEM with separate generators for distillation and CEM: 1) a faster EPR pair generator for distillation helps perform distillation rounds more frequently, but distillation failures also happen more frequently. 2) Even when distillation succeeds, the distilled pairs may decohere while waiting for the slower CEM formation. These insights show that both timing and resource balance critically affect the overall rate–fidelity tradeoff.

\noindent{\bf{Our Solution:}} We propose \methodName{}, a scheduler for the order of distillation and CEM formation and EPR pair generation allocation.  Unlike Static Distill+CEM, it does not commit to a fixed sequence; instead, it schedules distillation and CEM formation based on network demand. And, based on the insight that the request arrival time implicitly determines whether CEM is needed, \methodName{} operates {\em preemptively} (Figure~\ref{fig:intro_fig}(d)): if a request arrives during any step, the process stops, and the request is immediately served using the shared EPR pair.

\methodName{} scheduler selects from three priority policies that determine resource allocation and the order of distillation and CEM formation. {\em \approachZero{} (Distillation-First)} generates EPR pairs for distillation and then forms CEM, thus prioritizing service rate. It starts forming CEM after the successful completion of distillation process. {\em \approachOne{} (Memory-First)} first forms CEM, and then generates EPR pairs for distillation, thus prioritizing fidelity, as the distilled EPR pair will have the CEM ready for preservation. {\em \approachTwo{} (Concurrent)} concurrently distills EPR pairs and forms CEM, balancing between service rate and fidelity, and also applies with separate distillation and CEM formation EPR pair generators.

Our evaluations show that \methodName{} scheduler closes up to $11.21\%$ of the gap between ideal fidelity (unit) and the fidelity achieved by CED, and up to $49.48\%$
of the gap between the ideal service rate (equal to request rate) and the service rate of Static Distill+CEM.

\noindent{\bf Contributions:} Our paper's main contributions are:

\noindent\hangindent=1em\hangafter=0 $\bullet$ We introduce \methodName{}, a preemptive scheduler at node level for quantum networks that allocates EPR pair generation resources and schedules distillation, {\em Coding-Enhanced Memory}, and serving based on network demand.

\noindent\hangindent=1em\hangafter=0 $\bullet$ We propose three scheduling policies (Distillation-first, Memory-first, and Concurrent) optimized to increase fidelity, increase service rate, or balance both. 

\noindent\hangindent=1em\hangafter=0 $\bullet$ We show that \methodName{} improves the service 
rate (at the highest fidelity) and improves fidelity (at the highest service 
rate) compared to the state-of-the-art prior work.

%% file: sections/2_background.tex
\section{Background and Motivation }
\subsection{Quantum Network and Entanglement Distribution}\label{sec:QN_ED}

Nodes in a quantum network are connected via {\em entanglement distribution}, typically  {\em Einstein–Podolsky–Rosen (EPR) pairs}, through quantum channels such as optical fiber~\cite{wehner2018quantum}. Quantum networks enable diverse applications, e.g. distributed quantum computing~\cite{shapourian2025architectures, wu2022autocomm, wu2023qucomm},  quantum key distribution (QKD)~\cite{bennett2014quantum,ekert1991quantum}, and quantum sensing~\cite{qian2019heisenberg,qian2021optimal,eldredge2018optimal,ge2018distributed}. 

\begin{center} \textit{This work focuses on applications like QKD and sensing~\cite{wehner2018quantum, nist_quantum_networks}, where sustaining high service rates can be as critical as fidelity for the feasibility of applications.}
\end{center}

 When a sender node wants to distribute an EPR pair with a receiver node, the sender creates an EPR pair, keeps one qubit of the pair locally, and sends the other qubit of the pair to the receiver. This establishes a {\em link-level entanglement}, where two nodes directly share an EPR pair. While entanglement can also be extended to distant nodes through intermediate nodes using {\em entanglement swapping}~\cite{bennett1993teleporting,zukowski1993event}, this work focuses on direct link-level entanglement.

\begin{figure}[h!]
    \centering  \includegraphics[width=0.9\linewidth]{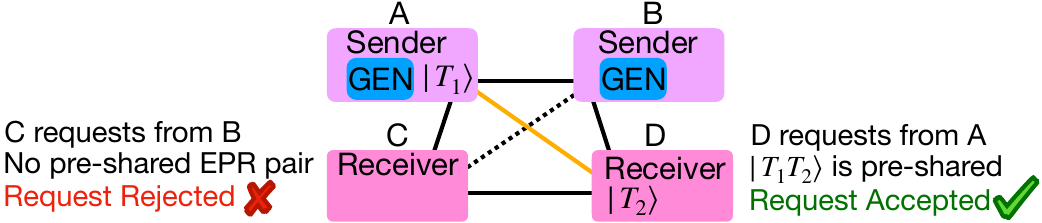}
        \caption{Receiver in quantum network requests to connect to Sender. If Sender has shared an EPR pair $\ket{T_1T_2}$ with Receiver the request is accepted; otherwise, rejected.}
    \label{fig:background_QN}
\end{figure}

Two typical distribution categories are~\cite{chakraborty2019distributed,dupuy2023survey}: {\em on-demand distribution} and {\em proactive distribution}. In on-demand distribution, two nodes start sharing an EPR pair only when one node  requests it from the other. Then, the distributed EPR pair is consumed immediately after preparation. This prevents decoherence impact on the fidelity of EPR pairs, but introduces a high service delay (low service rate). In contrast, in proactive distribution, EPR pairs are pre-shared regardless of the arrival of requests, eliminating request-to-service delay. When Receiver wants to use an EPR pair from Sender, the request is immediately accepted if a pre-shared EPR pair (e.g. $\ket{T_1T_2}$) exists; otherwise, it is rejected (Figure~\ref{fig:background_QN}). Thus, in proactive distribution, requests are either served or rejected upon arrival, {\em eliminating burst-induced delays but exposing EPR pairs to decoherence.} This work focuses on proactive distribution and reducing its associated decoherence impact.

\subsection{Entanglement Distillation}\label{sec:background:ED}
\begin{figure}[t!]
    \centering
        \includegraphics[width=0.9\linewidth]{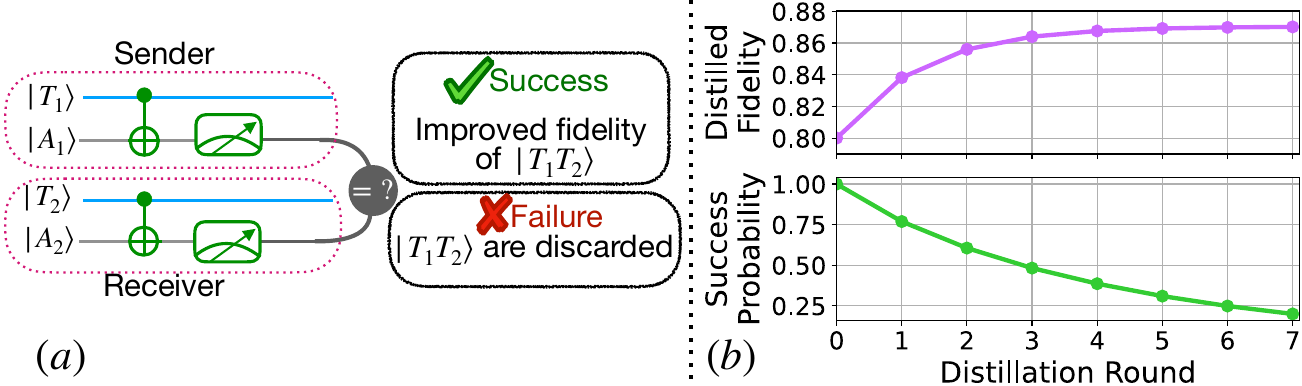} 
        \caption{(a) Entanglement Distillation. Equal results of measurement of $\ket{A_1A_2}$: success, improved fidelity of $\ket{T_1T_2}$. Unequal results: failure, corruption of both $\ket{A_1A_2}$ and $\ket{T_1T_2}$. (b) Fidelity and success probability across distillation rounds.}
    \label{fig:background_ED}
\end{figure}

The performance of the applications that use quantum networks depends on the fidelity of distributed entanglement. However, this fidelity is impacted by errors caused by different sources, such as depolarizing errors, imperfect EPR pair generators, and decoherence errors.
 {\em Entanglement distillation} improves fidelity by consuming multiple lower-fidelity EPR pairs to produce fewer, higher-fidelity pairs. For example, in Figure~\ref{fig:background_ED}(a), Sender and Receiver employ a simple {\em 2-to-1 protocol}~\cite{deutsch1996quantum, bennett1996purification, dur1999quantum, chitambar2014everything}, share a total of two entangled pairs, a target $\ket{T_1T_2}$ and an auxiliary $\ket{A_1A_2}$ pair to improve the fidelity of $\ket{T_1T_2}$. Using {\em local operations and classical communication (LOCC)}~\cite{chitambar2014everything}, they apply a local CNOT on their qubits, measure the auxiliary pair, and exchange results over a classical channel. If the results match, the fidelity of $\ket{T_1T_2}$ increases; if they mismatch, both pairs are discarded.
 
Figure~\ref{fig:background_ED}(b) shows the fidelity and success probability as a function of number of rounds of distillation using the {\em  Bilocal Clifford Protocol}~\cite{deutsch1996quantum, davies2024entanglement,kalb2017entanglement}.
 Let $F$ denote the fidelity of the target EPR pair that we aim to distill, which may have been distilled before or not. Let $F_{\text{aux}}$ denote fidelity of auxiliary EPR pair that is consumed (measured) in the distillation process. The fidelity of the target EPR pair, initially  $F$, improves to about $(1/3)F+(1+F_{\text{aux}})/3$ after a successful distillation round with probability of about $0.75$. Achieving $N$ successful rounds of distillation occurs with probability approximately $0.75^N$. Precise general-case equations for fidelity improvement and success probability are presented in Section~\ref{subsec:evalMethods_distillation}. As shown in Figure~\ref{fig:background_ED}(b), starting from $F = 0.8$, the maximum achievable fidelity is about $0.87$, but with a low probability of $0.237$.

\subsection{Rate-Fidelity Tradeoffs in Prior Work Policies}
\noindent\textbf{{Continuous Entanglement Distillation (CED)}} CED policy~\cite{davies2024entanglement, elsayed2023fidelity, mundarain2009entanglement} continuously allocates EPR pair resources to share auxiliary EPR pairs and distill the target EPR pair until the target pair is consumed. This policy is particularly beneficial for proactive distribution, where EPR pairs are not immediately consumed and are, therefore, more susceptible to decoherence over time. However, CED reduces the number of EPR pairs usable for serving requests in two main ways: 1)~Continuously performing distillation rounds consumes many auxiliary EPR pairs to produce a single high-fidelity target pair, reducing the available rate for serving requests. 2)~Unsuccessful distillation destroys the target EPR pair along with all the previously measured auxiliary pairs, without getting any benefit. Since success probability decreases exponentially with each round (Figure~\ref{fig:background_ED}(b)), CED's static resource allocation policy of ``always distill" mitigates decoherence (higher fidelity) but causes substantial rate loss, especially under low EPR pair consumption demand.

\begin{figure}[t!]
\centering
\includegraphics[width=0.8\linewidth]{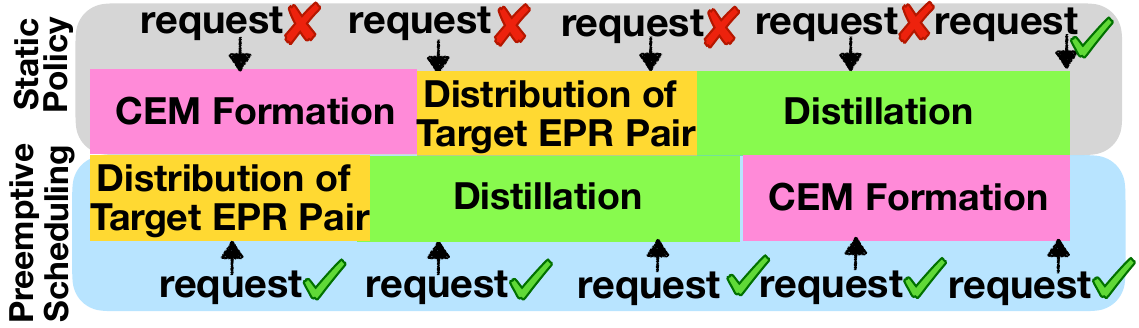} 
\caption{Static vs. Preemptive Scheduling. Unlike Static policies, preemptive demand-aware scheduling adapts distillation/CEM order and serves requests at any stage.}
\label{fig:StaticVsDynamic}
\end{figure}

\noindent\textbf{{CEM and Its Combination with Distillation}}\label{sec:coding+distillation}
\new{Another policy to mitigate errors is using quantum error correction to form fault-tolerant qubits, what we refer to as {\em Coding-Enhanced Memory (CEM)}. We call a design that exclusively uses CEM to maintain fidelity as {\em Static CEM}. 
CEM can also be combined with distillation, to increase and then maintain fidelity~\cite{muralidharan2016optimal, patil2024entanglement, pathumsoot2024boosting}, what we refer to as  {\em Static Distill+CEM} policy. Static Distill+CEM follows a fixed order: first CEM formation, then distillation and preservation. These static policies commit to predetermined sequences regardless of request arrival. EPR pairs become available only after completing the entire fixed pipeline (Figure~\ref{fig:StaticVsDynamic}). This rigid ordering and the inability to adapt the order of techniques create suboptimal rate-fidelity tradeoffs. Static Distill+CEM achieves high fidelity but suffers severe rate reduction from the lengthy sequential process. Instead, a scheduler that can preemptively serve requests and dynamically order distillation and CEM can improve the rate-fidelity tradeoff. Prior policies are
summarized in Table~\ref{tab:comparison}.}

\begin{table}[h!]
\centering

\caption{Comparison of Quantum Network Policies}
\renewcommand{\arraystretch}{0.95} 
\setlength{\tabcolsep}{0.1cm} 
\begin{tabular}{|m{0.44\columnwidth}|>{\centering\arraybackslash}m{0.26\columnwidth}|>{\centering\arraybackslash}m{0.18\columnwidth}|}
\hline
\textbf{Policy} & \textbf{Method} & \textbf{Serving} \\ \hline
CED~\cite{davies2024entanglement, elsayed2023fidelity, mundarain2009entanglement} & Fixed\newline(Only Distill) & Reject interim \\ \hline
Static (Distill+)CEM~\cite{muralidharan2016optimal,jiang2009quantum,fowler2010surface,muralidharan2014ultrafast,munro2012quantum, patil2024entanglement, pathumsoot2024boosting}& Fixed (CEM($\rightarrow$Distill)) & Reject interim  \\ \hline
Our Proposal  & Demand-aware scheduling & Preemptive \\ \hline
\end{tabular}
\label{tab:comparison}
\end{table}

\vspace*{-3mm}
\subsection{EPR Pair Generation Resource}

\new{\noindent\textbf{Shared EPR pair generator}  Depending on qubit modality, networking and local computation may share entanglement sources, which must time-multiplex between generating EPR pairs for remote communication and for local operations such as error correction.
Examples include a squeezed-state generator in photonic systems such as Xanadu's
Aurora~\cite{aghaee2025scaling}, a spontaneous four-wave mixing microring in silicon
photonics~\cite{llewellyn2020chip}, and an NV-center electron spin in solid-state
nodes~\cite{pompili2021realization, hermans2022qubit}. Uncoordinated access starves either the network, when local tasks monopolize the generator, or computation, when network requests block local progress.}

\noindent\textbf{\new{Dedicated EPR pair generator:}} In some qubit modalities, networking and
local computation rely on physically separate entanglement sources. For example, in
trapped-ion systems, dedicated network qubits (e.g., $^{88}$Sr$^+$) handle remote
entanglement generation via photonic links, while separate resources (e.g.,
$^{43}$Ca$^+$) perform local computation~\cite{main2025distributed}. Similarly,
superconducting architectures provision distinct transmon qubits for local gates and
separate transduction modules for inter-node networking~\cite{almanakly2025deterministic}.
Scheduling is still necessary: timing dependencies between distillation, CEM formation, and consumption mean naive resource scaling does not resolve contention.  Faster generators increase failure rates. Successfully distilled pairs may
decohere while waiting for downstream operations. These challenges are inherent to the
quantum network node, independent of whether the generation hardware is shared or
dedicated;{ \em this work considers both scenarios.}

\subsection{Figure-of-Merit}\label{sec:fig_merit}
{\em Fidelity} and {\em service rate} are critical metrics in quantum networks, with applications prioritizing one over the other depending on their needs. We use \textbf{{\em Fidelity Gap Improvement}}, defined as $(F_2 - F_1)/(1 - F_1) \times 100$, to show what percentage of the gap to ideal fidelity ($F=1$) is closed by improving the fidelity from $F_1$ to $F_2$.  

In proactive distribution, requests are served immediately if EPR pairs are pre-distributed; otherwise, rejected. To quantify this, we define the {\em normalized service rate} as $R_s / \mu$, where $R_s$ is service rate, the rate of successfully distributed EPR pairs available at the time of request (and accordingly consumed for serving the request), and $\mu$ is the consumption request rate. Ideally, if all requests are served, $R_s/\mu = 1$. We define \textbf{\em Rate Gap Improvement} as $(R_{s_2}/\mu - R_{s_1}/\mu)/(1 - R_{s_1}/\mu) \times 100$, showing how much of the gap to ideal rate is closed by changing $R_{s_1}$ to $R_{s_2}$.

\subsection{Goal}

\begin{figure}[h!]
\centering
\includegraphics[width=\linewidth]{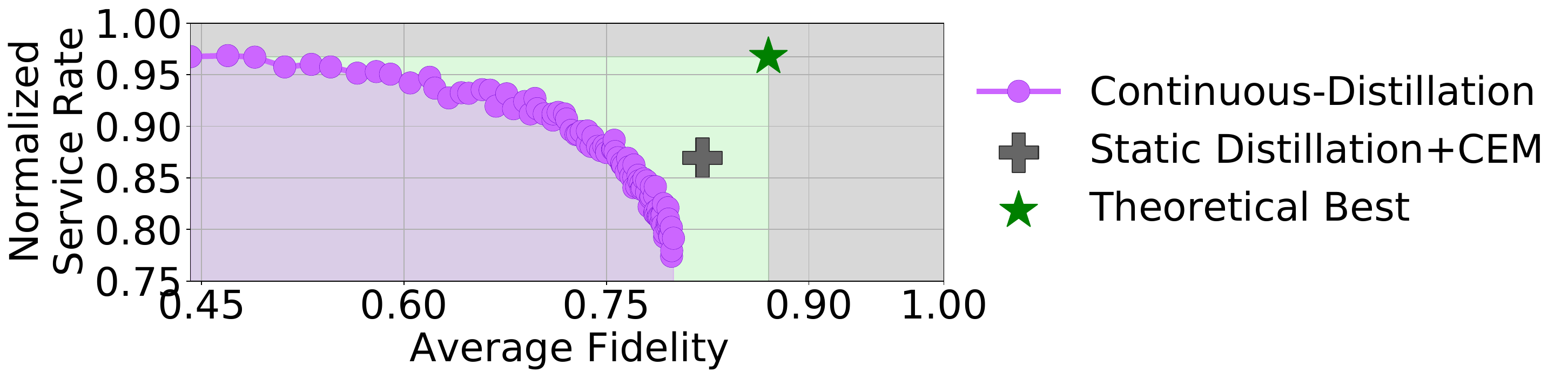} 
\caption{The rate-fidelity tradeoff in quantum networks. There exists a gap (green area) between prior work and theoretical best achievable service rate and fidelity.}
\label{fig:goal_hypothetical_best}
\end{figure}

Figure~\ref{fig:goal_hypothetical_best} illustrates the rate-fidelity tradeoff in quantum networks. Prior static policies achieve suboptimal tradeoffs due to fixed sequences and resource allocation that ignores network demand. The timing interactions between distillation and CEM formation mean that simply increasing resources does not guarantee better performance. Ideally, we aim for unit service rate and fidelity, but theoretical limits in current distillation protocols make perfect fidelity and rate practically unattainable (gray area in Figure~\ref{fig:goal_hypothetical_best}). Nonetheless, there exists a gap between prior work and the best achievable case (green area in Figure~\ref{fig:goal_hypothetical_best}).

\textit{Our objective is to close this gap and enable a better balance
for rate-fidelity tradeoff through demand-aware scheduling of distillation and CEM formation at the node level.} This scheduling aims to increase the area under the rate-fidelity curve, pushing closer to the ideal scenario and enhancing the efficiency of quantum networks.

%% file: sections/3_evaluation_methodology.tex
\section{Evaluation Methodology}\label{Sec:Methodology}
For evaluating \methodName{}, we implement the simulation based on Continuous-Time Markov Chain (CTMC), which is the standard framework for modeling quantum networks, including the baseline in prior work~\cite{davies2024entanglement}, and many other prior works such as~\cite{vardoyan2023capacity,shchukin2019waiting,nain2020analysis,chandra2022scheduling,vardoyan2022quantum} which model complex networks. Details of the models are provided in Appendix~\ref{apx:design-markov}.

\subsection{Entanglement State and Noise Model}\label{sec:noise_model}
\noindent\textbf{Depolarizing Error and Imperfect EPR Pairs:} We capture the effect of errors on EPR pairs by using a {\em Werner state}, a widely accepted approach in quantum networks and quantum information theory~\cite{werner1989quantum,bennett1996mixed, horodecki1997separability, davies2024entanglement}. Equation~\ref{eq:Werner_states} denotes the Werner state, capturing depolarization noise, a dominant error in entanglement distribution.   $\{\ket{\phi^+}, \ket{\psi^+}, \ket{\psi^-}, \ket{\phi^-} \}$ are Bell states. The fidelity of our desired state, represented by $\ket{\phi^+}$, is quantified by the parameter $F$.

\begin{equation}\label{eq:Werner_states}
    \begin{aligned}
        \small\rho_w = F\ket{\phi^+}\bra{\phi^+} + \frac{1-F}{3} ( \mathbb{I} - \ket{\phi^+}\bra{\phi^+} )
    \end{aligned}
\end{equation}

Depolarizing noise changes the input state into a maximally mixed state, theoretically the most difficult to correct~\cite{bennett1996mixed, horodecki1997separability}. 
We test our models across different F values, allowing us to assess the performance under varying noise levels. 

\noindent
\textbf{Decoherence Error:}
For Werner states, there exist well-established analytical models that describe how decoherence degrades fidelity over time. The impact of decoherence on the fidelity $F(t)$ of Werner state after $\Delta t$ is denoted in Equation~\ref{eq:phys_decoh_f}~\cite{dur2005standard}, where the memory lifetime is $1/\Gamma$ ($\Gamma$ is the decoherence rate). In our experiments, we set $\Gamma$ of physical qubits to $0.05$,  within the range in baseline~\cite{davies2024entanglement}. We simulate CEM with lower $\Gamma$ values. 

    \begin{equation}\label{eq:phys_decoh_f}
        F(t+\Delta t) = e^{- \Gamma \Delta t} ( F(t) -\frac{1}{4})+\frac{1}{4}
    \end{equation}

\subsection{Attenuation}\label{sec:methodology_atten}
The success probability of distributing a link-level EPR is $p = 10^{-\alpha l}$, where we set $l = 1km$, the distance between nodes, and $\alpha= 0.02$ is the attenuation rate in optical fiber~\cite{pouryousef2024resource}. The transmission time is considered for decoherence error.

\subsection{ Distillation Protocol}~\label{subsec:evalMethods_distillation}
We use the 2-to-1 Bilocal Clifford protocol~\cite{deutsch1996quantum, davies2024entanglement}, whose success probability and distilled fidelity are given by  Equations~\ref{eq:ev_p_jump_function_intro} and~\ref{eq:ev_f_jump_function_intro}. Therein, $F$  represents the fidelity of the target entangled state we aim to improve, which may have been distilled before or not. $F_{\text{aux}}$ corresponds to the fidelity of auxiliary shared EPR pair consumed (measured) during the distillation process. The target EPR pair’s fidelity $F$ improves to $F_{\text{distilled}}$ with success probability $P_{\text{success}}$.

\begin{equation}\label{eq:ev_p_jump_function_intro}
    P_{\text{success}}= \frac{1}{9}({(1+2F)(1+2F_{\text{aux}})+4(1-F)(1-F_{\text{aux}})})
\end{equation}
\vspace*{-0.05in}\begin{equation}\label{eq:ev_f_jump_function_intro}
    F_{\text{distilled}}= \frac{9FF_{\text{aux}}+(1-F)(1-F_{\text{aux}})}{9P_{\text{success}}}
\end{equation}

%% file: sections/4_model.tex
\section{\methodName{}: Key Insights}
\noindent\textbf{Temporal Dependency of CEM Utility:} The benefit of CEM depends critically on the duration between successful distillation and request arrival. When this duration is short, decoherence impact is minimal and CEM overhead is not justified (recently distilled EPR pairs can serve requests with acceptable fidelity). Conversely, when the duration between successful distillation and request arrival is long, qubits become vulnerable to decoherence, making CEM essential for maintaining high fidelity.

\noindent\textbf{Resource Scaling Limitations:} Simply adding more resources does not solve the fundamental timing challenges. For example, faster EPR pair generator dedicated to distillation increases distillation frequency but also increases failure rates. Even when distillation succeeds, the distilled pairs may decohere while waiting for slower CEM formation. Therefore, these timing dependencies require scheduling rather than naive resource scaling.

\noindent\textbf{Request Arrival as Implicit Signal:} Request arrival timing implicitly determines an efficient strategy. Early arrivals indicate that CEM overhead should be avoided (serve immediately using physical qubits). Late arrivals suggest sufficient time exists for CEM formation. This insight motivates {\em preemptive} serving: rather than committing to fixed sequences, the system should serve requests during intermediate stages and let arrival patterns guide resource allocation decisions.

\noindent\textbf{Demand-Aware Scheduling:} Different network applications require different scheduling policies. High-rate demand scenarios benefit from prioritizing service rate by minimizing CEM usage, while low-demand periods can use available idle time to form CEM and serve with higher fidelity. No node-level scheduler currently exists that adapts technique ordering and resource allocation based on the demands of applications running on the network.

Based on the above insights, we propose three policies in \methodName{} to optimize for different fidelity and rate requirements: 1)~{\em Distillation-first}: \approachZero{} prioritizes distribution and distillation of EPR pairs to achieve a higher service rate. 2)~{\em Memory-first}: \approachOne{} prioritizes fidelity and \LogicalQ{} formation as a proactive approach to ensure that a \LogicalQ{} is immediately available to preserve a high-fidelity EPR pair as soon as it is distilled. 3)~{\em Concurrent}: \approachTwo{} balances both fidelity and rate by distilling and forming CEM concurrently, then shifts to only distillation once
CEM is formed, also useful when multiple EPR generator resources are available.

\begin{figure*}[ht]
    \centering
    \includegraphics[width=0.9\linewidth]{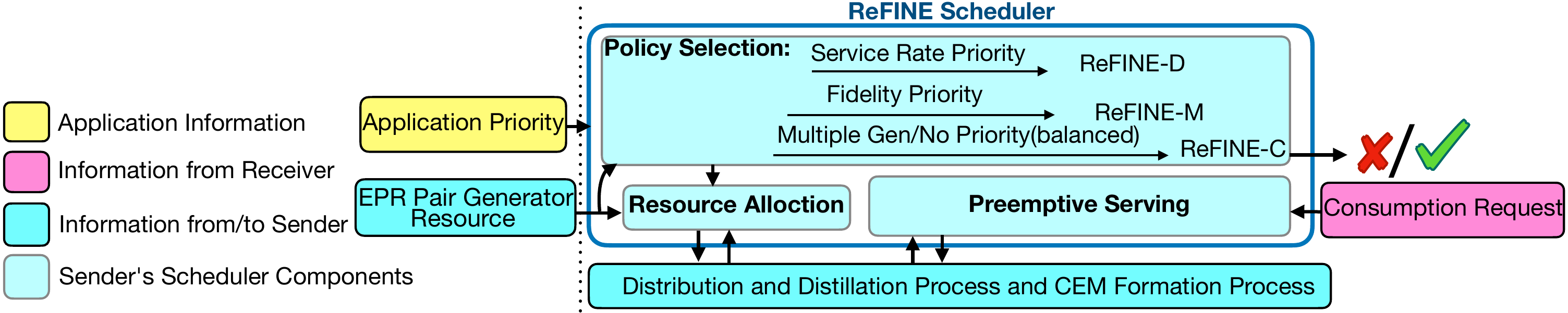}
    \caption{\methodName{} Overview: The scheduler selects policies based on application requirements, allocates EPR generator resources, and serves requests preemptively during distillation and CEM formation processes.}
    \label{fig:overview}
\end{figure*}

\section{\methodName{} Design}\label{sec:Refine Design}

\methodName{} is a node-level scheduler deployed at Sender's side.  Sender and Receiver connect via a quantum channel; Sender owns EPR pair generators for entanglement distribution and CEM formation, and sends one qubit of an EPR pair to Receiver, during the distribution.

    \methodName{} focuses on {\em communication qubits}, which are responsible for establishing entanglement between remote nodes, independent of whether the computation qubits are fault-tolerant or NISQ-based~\cite{kimble2008quantum, wehner2018quantum, briegel1998quantum}. For applications like QKD, where shared EPR pairs are measured upon request, computation only consists of measurement, that can be performed on either physical or logical qubits~\cite{ muralidharan2016optimal, wehner2018quantum}. For applications requiring operations on shared EPR pairs, the communication qubit’s data is transferred to computation qubits, which may be either logical or physical, depending on the fault tolerance level of the system~\cite{pattison2025constant}.

\subsection{Sender Node Architecture} \label{sec:Alice_arc}
Sender is equipped with the following: 

\noindent{\bf{Physical qubits}}: 
    For distillation, Sender and Receiver share two entangled pairs, target EPR pair and auxiliary EPR pair. Sender uses two physical qubits to store half of each pair. During distillation, the auxiliary pair is measured, and the process is successful if the measurement results match, which increases the fidelity of the target EPR pair. Decoherence reduces the fidelity of entangled pairs shared between the Sender and Receiver.
    
\begin{figure}[h!]
    \centering
    \includegraphics[width=0.5\columnwidth]{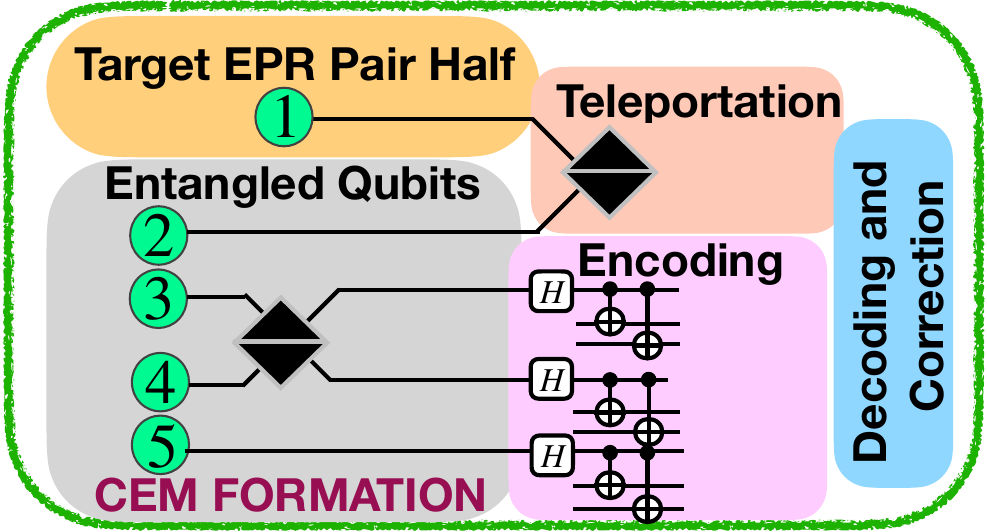}
     \caption{Coding Enhanced Memory (CEM). CEM requires two EPR pairs, namely (\textcircled{2},\textcircled{3}) and (\textcircled{4},\textcircled{5}), and uses teleportation for initialization~\cite{luo2021quantum}.}
    \label{fig:CEM}
    
\end{figure}

\noindent{\textbf{Coding Enhanced Memory (CEM)}}: This memory, equip\-ped with QEC, forms a \LogicalQ{} for reliable, decoherence-tolerant storage of one half of a distilled (target) EPR pair, with the other half shared with Receiver. Inspired by Luo et al.~\cite{luo2021quantum}, we use 9-qubit Shor’s code and teleportation for initialization, while noting that \methodName{} naturally extends to other error-correcting codes with appropriate modifications. In Figure~\ref{fig:CEM}, qubit \textcircled{1} is half of the target EPR pair stored as a physical qubit in Sender's side. Qubits \textcircled{2} and \textcircled{4} are entangled with qubits \textcircled{3} and \textcircled{5}, respectively, provided by the EPR pair generator. First, a Bell State Measurement (BSM) between qubits \textcircled{3} and \textcircled{4} creates the Greenberger-Horne-Zeilinger ($\text{GHZ}_4$) state. We refer to these steps as \textbf{{\em forming CEM}}. Then, {\em \bf{CEM initialization}} involves another BSM between qubits \textcircled{1} and \textcircled{2} teleports the data of qubit \textcircled{1} into \LogicalQ{}. This preserves one qubit of the distributed target EPR pair in \LogicalQ{}, maintaining entanglement between Sender’s \LogicalQ{} and Receiver’s other half of the target EPR pair. 

 Note that the preparation of the $\text{GHZ}_4$ state (CEM formation) is independent of the state of qubit \textcircled{1}. Thus, the necessary steps to create the required $\text{GHZ}_4$ state—consisting of two EPR pairs and a BSM operator—can be performed before, after, or concurrently with the preparation of qubit \textcircled{1}~\cite{luo2021quantum}. This flexibility ensures that forming CEM prior to distillation is not only feasible but also functionally equivalent to forming it afterward. 

\noindent{\textbf{EPR Pair Generator}}: Sender uses either one or two generators. With one, each EPR pair is either shared with the Receiver or used locally for CEM formation. With two, one handles distribution while the other supports local CEM formation.

\begin{figure*}[t]
    \centering
    \includegraphics[width=0.955\linewidth]{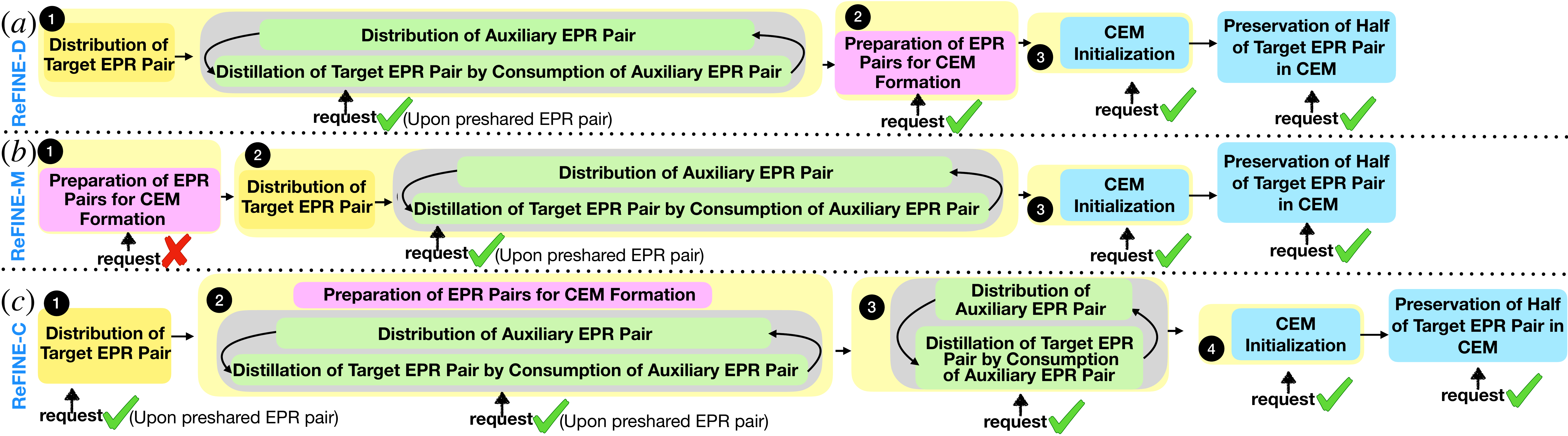}
    \caption{ Overview of \methodName{} Scheduling and its preemptive serving. (a) Distillation-first (b) Memory-first (c) Concurrent.}
    \label{fig:policies}
\end{figure*}
\subsection{Receiver Node Architecture} 
Similar to Sender, Receiver uses two physical qubits to store its halves of the target and auxiliary EPR pairs. After receiving an EPR half in a physical qubit, Receiver may use any coding scheme, or none, independently of the Sender’s scheme. We assume it preserves high-fidelity EPR pairs using coding.

\subsection{\approachZero{}: Distillation-First Policy}\label{sec:label-first}
This scheduling policy prioritizes service rate by allocating EPR generator resources first to distillation operations, with CEM formation scheduled only after distillation completion. The scheduler executes the following resource allocation sequence as illustrated in Figure~\ref{fig:policies}(a): \noindent\textbf{\textcircled{1} Distillation Step:} The scheduler allocates all EPR generator capacity to distillation operations. Sender shares the target EPR pair with Receiver, then generates auxiliary EPR pairs for each distillation round to improve the shared target's fidelity. In this step, half of the shared EPR pairs are stored in the physical qubits until a sufficient number of successful distillation rounds has been applied to the target EPR pair.
\noindent\textbf{\textcircled{2} CEM Formation Step:} Once distillation rounds complete, the scheduler reallocates EPR generator resources from distillation to \LogicalQ{} formation, dedicating all subsequent generation capacity to CEM creation.
\noindent\textbf{\textcircled{3} Preservation Step:} After \LogicalQ{} formation, \approachZero{} schedules CEM initialization to preserve the high-fidelity target EPR pair in Sender's \LogicalQ{}. Note that during the time that \approachZero{} creates the \LogicalQ{}, the distilled EPR pair is impacted by decoherence. Therefore, the fidelity of the preserved EPR pair is lower than the fidelity achieved after the completion of distillation rounds.

\noindent\textbf{Preemptive Serving:} \approachZero{} serves requests immediately upon arrival, regardless of processing stage, if a preshared EPR pair exists. In step \textcircled{1}, the scheduler serves from physical qubits; in steps \textcircled{2} and \textcircled{3}, it serves using distilled EPR pair from physical qubits and from CEM, respectively. This preemptive scheduling eliminates request rejections while EPR pairs await processing completion.

\subsection{\approachOne{}: Memory-First Policy}\label{sec:memory-first}

This scheduling policy prioritizes fidelity by allocating EPR generator resources first to CEM formation, ensuring immediate preservation ability once distillation completes. Comparing with \approachZero{}, \approachOne{} reverses the resource allocation sequence, executing CEM formation before distillation operations, shown in Figure~\ref{fig:policies}(b).

The key scheduling difference is the {\em fidelity-optimized resource allocation}, that \approachOne{} begins by allocating EPR generator capacity to \LogicalQ{} formation, with no shared EPR pairs available initially. Investing in CEM formation upfront enables immediate preservation of distilled pairs without decoherence, maximizing fidelity.

\noindent\textbf{Preemptive Serving:} If requests arrive after CEM formation but before CEM initialization, the scheduler serves from physical qubits rather than rejecting them. The initial request rejections during early CEM formation represent the cost of prioritizing fidelity, but \approachOne{}  minimizes this impact through preemptive serving as soon as CEM formation completes and an EPR pair get preshared.

\subsection{\approachTwo{}: A Concurrent Scheduling Policy}\label{sec:concurrent-policy}

This scheduling policy balances fidelity and service rate by concurrently allocating EPR generator resources to both distillation and CEM formation. This policy is also adaptable with multiple EPR generators, enabling dedicated resource allocation to each process. The scheduler executes the following concurrent resource allocation sequence as illustrated in Figure~\ref{fig:policies}(c):
\noindent\textbf{\textcircled{1} Initial Serving Setup:} The scheduler immediately shares a target EPR pair with Receiver to ensure immediate request serving capability from the start.
\noindent\textbf{\textcircled{2} Concurrent Resource Step:} The scheduler divides EPR generation capacity based on consumption rate $\mu$: allocating $\mu$ to distillation operations while dedicating the rest to CEM formation. This concurrent scheduling allows both distillation and CEM formation to progress simultaneously. The scheduler strategically delays the final distillation round until CEM formation completes, ensuring fully distilled EPR pairs can be preserved immediately and avoid further decoherence.
\noindent\textbf{\textcircled{3} Completion and Reallocation Step:} Once CEM formation completes and the final distillation round executes, the scheduler reallocates all EPR generation capacity exclusively to distillation operations for maximum throughput.

\noindent\textbf{Preemptive Serving:} \approachTwo{} serves requests immediately in all stages, from physical qubits initially, then from preserved CEM pairs when available.

%% file: sections/5_evaluations.tex
\section{Evaluation}
We evaluate \methodName{} with generation rate $\lambda = 1$ and three $\mu$ values representing low, moderate, and high consumption rates.  Initial fidelity is set to $F_0 = 0.8$, according to baseline~\cite{davies2024entanglement}, and we provide  sensitivity analysis on $F$ in Section~\ref{sec:initf_s}. Note that static baselines offer only a single tradeoff point. Section~\ref{sec:MG} evaluates \methodName{} with separate EPR generators for distillation and CEM formation, while other sections use a shared generator. We set $\Gamma = 0.05$ and $5\times 10^{-4}$ for physical qubits and CEM, respectively. We provide sensitivity analysis to $\Gamma$ values in Section~\ref{apx:gammaSen}.

\subsection{Impact on the Rate-Fidelity Tradeoff}\label{sec:tradeoff}

\begin{figure*}[t]
    \centering
    \includegraphics[width=\linewidth]{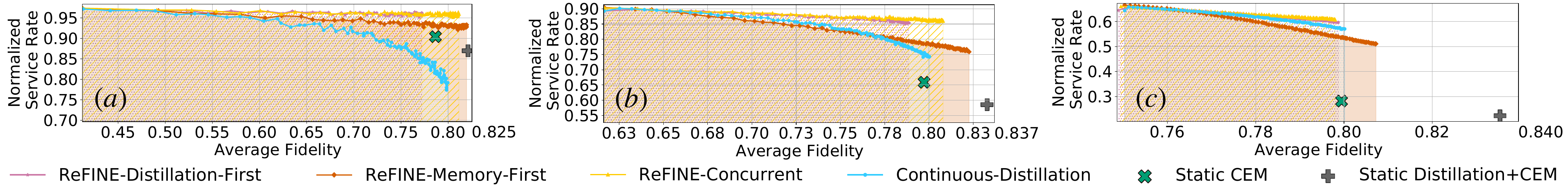}
    \caption{Rate-fidelity tradeoff across consumption regimes. (a)~$\mu = 0.02\lambda$ (low): \approachOne{} matches the fidelity of Static Distill+CEM at higher rate. (b)~$\mu = 0.1\lambda$ (moderate): \approachTwo{} achieves the best balance up to its maximum achieved fidelity. (c)~$\mu = 0.5\lambda$ (high): for the common achievable fidelity, \approachZero{} offers the best overall tradeoff. \emph{No single fixed-order policy wins across all three regimes.}}
    \label{fig:tradeoff}
\end{figure*}

\noindent{\bf{Low Consumption Rate:}}\label{l_consum_results}
Figure~\ref{fig:tradeoff}(a) shows the tradeoff between service rate and average fidelity for a low consumption rate of $\mu=0.02\lambda$.  \methodName{} achieves a better rate-fidelity tradeoff with $48.66\%$ Rate Gap Improvement, over Static Distill+CEM while maintaining comparable fidelity ($-0.85\%$ gap). \approachOne{} also outperforms CED gains $9.98\%$ and $67.86\%$ Fidelity and Rate Gap Improvement, respectively.

\textit{{Why \approachOne{} excels:}} This regime favors \approachOne{} because the low consumption rate minimizes the impact of request rejections during CEM formation. With fewer requests arriving, \approachOne{} can prioritize forming CEM without significantly compromising the overall service rate.

\noindent{\bf{Moderate Consumption Rate:}} Figure~\ref{fig:tradeoff}(b) shows the rate-fidelity tradeoff for $\mu=0.1\lambda$. \approachTwo{} achieves the best balance with $66.26\%$ Rate Gap Improvement over Static Distill+CEM while maintaining $3.85\%$ Fidelity Gap Improvement over CED. \approachOne{} achieves $11.21\%$ and $6.26\%$ Fidelity and Rate Gap Improvement over CED.

\textit{Why \approachTwo{} excels:} The balanced request arrival pattern allows both distillation and CEM formation to proceed simultaneously without overwhelming either process. Unlike Static Distill+CEM's sequential bottleneck or \approachOne{}'s initial rejection period, \approachTwo{} can serve requests immediately while building both fidelity and preservation capabilities in parallel. This concurrent resource allocation matches the moderate demand timing perfectly and also works for the cases that priorities are unknown.

\noindent{\bf{High Consumption Rate:}} Figure~\ref{fig:tradeoff}(c) shows the rate-fidelity tradeoff for $\mu=0.5\lambda$. \approachZero{} and \approachTwo{} achieve similar performance to CED, with \approachZero{} providing slight Rate Gap Improvement over CED while maintaining comparable fidelity. \approachZero{} outperforms \approachTwo{} in this regime due to its distillation-first priority. For applications that need higher fidelity values, \approachOne{} and Static Distill+CEM offer $4.22\%$ and $18.18\%$ Fidelity Gap Improvement respectively, at the cost of $-17.56\%$ and $-48.15\%$ Rate Gap Improvement compared to \approachZero{}.

\textit{Why \approachZero{} excels:} This regime favors \approachZero{} because frequent request arrivals leave insufficient time for CEM formation while CEM usage is also unnecessary as EPR pairs are consumed rapidly with negligible decoherence impact. \approachZero{} dedicates all resources to immediate distillation and serving, matching the high-demand timing, while Static methods suffer severe rate penalties  under heavy load.

\textit{Adaptive Performance Across Demand Regimes:} \methodName{} demonstrates demand-aware adaptability that static policies cannot achieve. While static methods excel only in extreme regimes (Static Distill+CEM at very low rates, CED at very high rates), \methodName{}'s dynamic scheduling adapts resource allocation to match network demand timing, enabling a better, more efficient  balance for the rate-fidelity tradeoffs across all consumption patterns rather than committing to fixed sequential processing.

\subsection{Separate EPR Pair Generators}\label{sec:MG}
\begin{figure}[h]
    \centering
    \includegraphics[width=0.95\columnwidth]{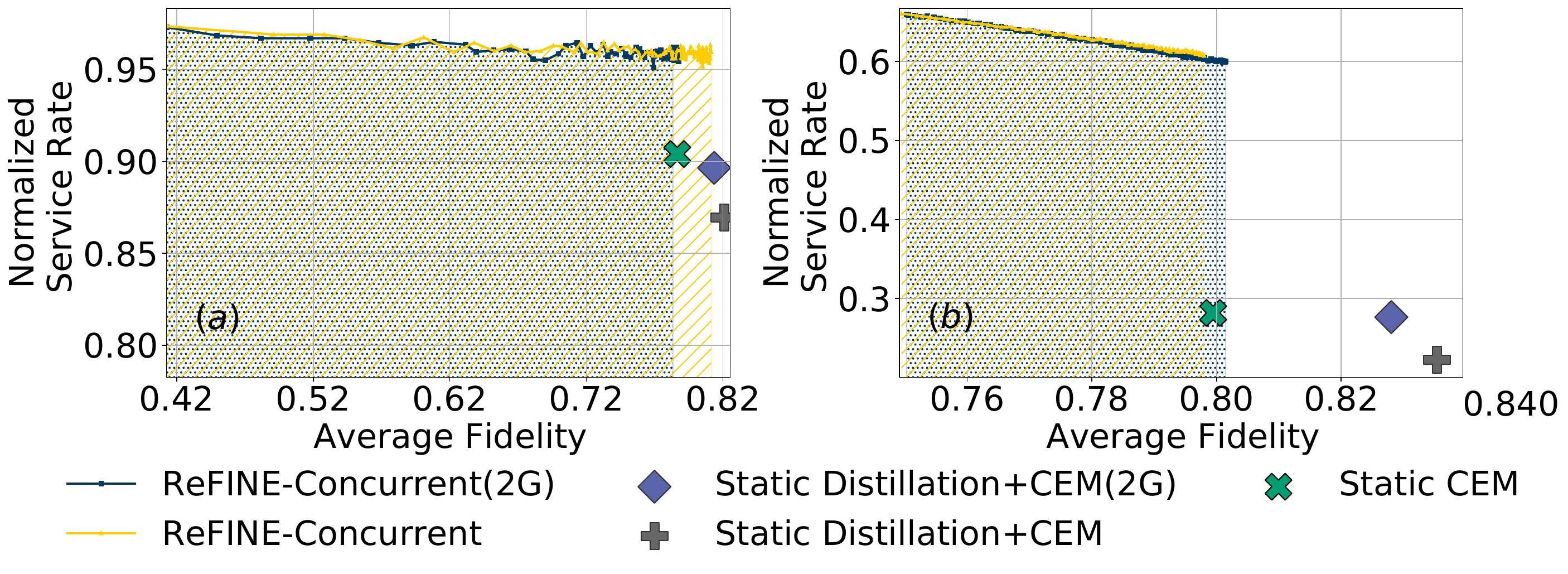}
    \caption{Rate-fidelity tradeoff (a) low consumption ($\mu = 0.02\lambda$) and (b) high consumption ($\mu = 0.5\lambda$); ``(2G)" denotes two separate generators, one for distillation and one for CEM formation. More resources does not help static policies.}
    \label{fig:multigen}
\end{figure}

Figure~\ref{fig:multigen} shows that Static Distill+CEM with separate generators achieves higher rate approaching Static CEM levels by parallelizing and eliminating the distillation time  bottleneck. However, fidelity cannot improve due to timing challenges; for example, distilled pairs may decohere while waiting for CEM formation, confirming that the fundamental problem is poor scheduling, not resource scarcity.
Unlike static policies, ReFINE can efficiently utilize additional resources when demand justifies it. Figure~\ref{fig:multigen}(b) shows ReFINE achieves slightly higher fidelity with multiple generators under high consumption rates, while at low consumption rates (Figure~\ref{fig:multigen}(a)), a single generator is more beneficial as it avoids unnecessary timing complications and decoherence effects. 

\new{\subsection{Policy Selection and Scope of Demand-Awareness}
ReFINE selects among its three policies according to the application's demand class, which is a provisioning-time property of the target workloads. QKD links, sensing arrays, and interferometric baselines are deployed for a characteristic, stable consumption regime rather than adversarial bursty traffic; the scheduler's task is therefore to map a known demand class to the policy that is optimal for it. We establish this mapping empirically (Figure~\ref{fig:tradeoff}): ReFINE-M for low, ReFINE-C for moderate, and ReFINE-D for high consumption rates. The mapping is non-obvious because resource interactions are non-monotonic: a faster generator increases distillation frequency but also failure rate, and distilled pairs decohere while awaiting CEM formation, so the demand-optimal policy is not predictable from resource counts alone.  Selecting the policy for a known class is an O(1) decision.}

\subsection{An Application: QKD Fidelity and Rate}\label{sec:QKD}
\begin{figure}[h]
    \centering
    \includegraphics[width=0.8\columnwidth]{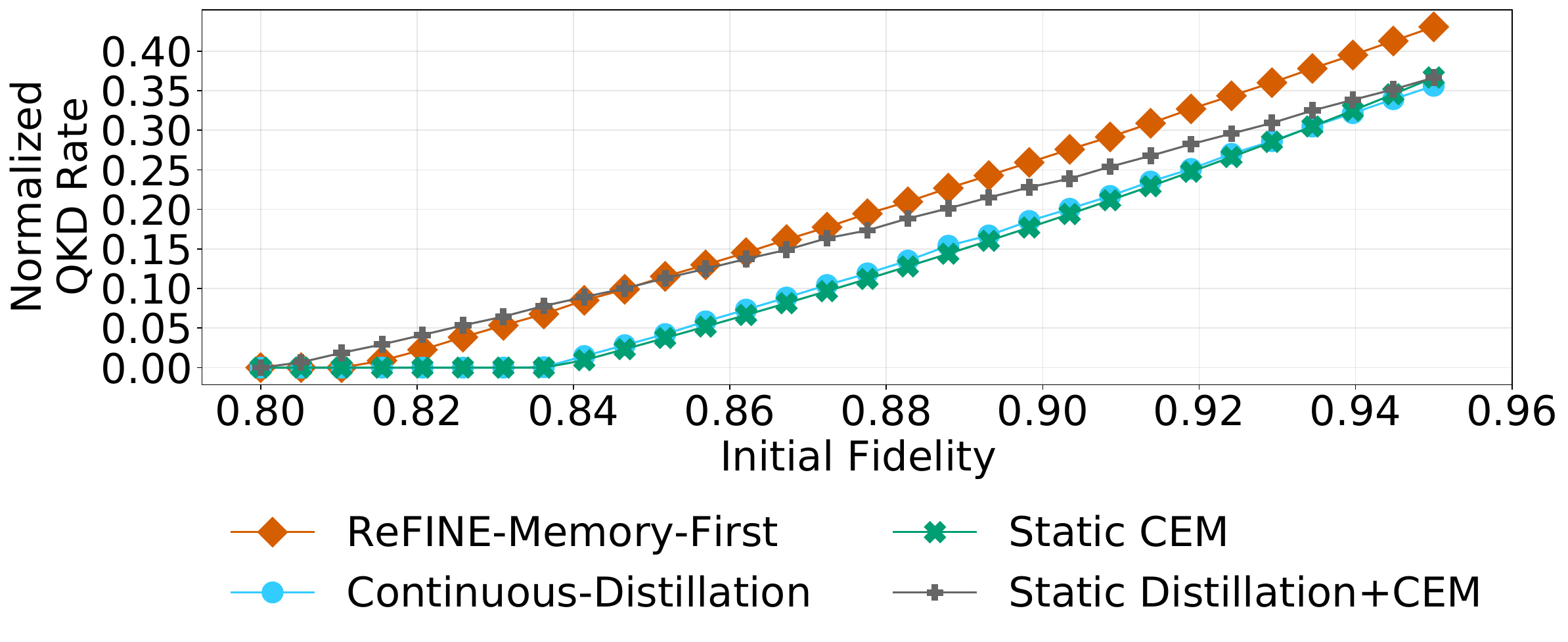}
    \caption{QKD rate for different initial fidelity. \approachOne{} prioritizes fidelity and achieves the highest rate, boosting more pairs above the fidelity threshold.}
    \label{fig:QKD_rate}
\end{figure}

We simulate the entanglement distribution rate for QKD by varying the initial fidelity of shared entangled pairs, considering the $0.835$ threshold, fidelity based on prior work~\cite{gottesman2004security,scarani2009security}, for $\mu = 0.1\lambda$. Our results confirm that \approachOne{}, which prioritizes fidelity, achieves a greater fidelity boost, \approachOne{} ensures more EPR pairs surpass the threshold, resulting in a higher effective QKD rate, closing the gap by $10.13\%$, as shown in Figure~\ref{fig:QKD_rate}.
\subsection{Sensitivity Analysis }\label{sec:initf_s}
\subsubsection{Initial Fidelity}
\begin{figure}[h!]
    \centering
    \includegraphics[width=0.95\columnwidth]{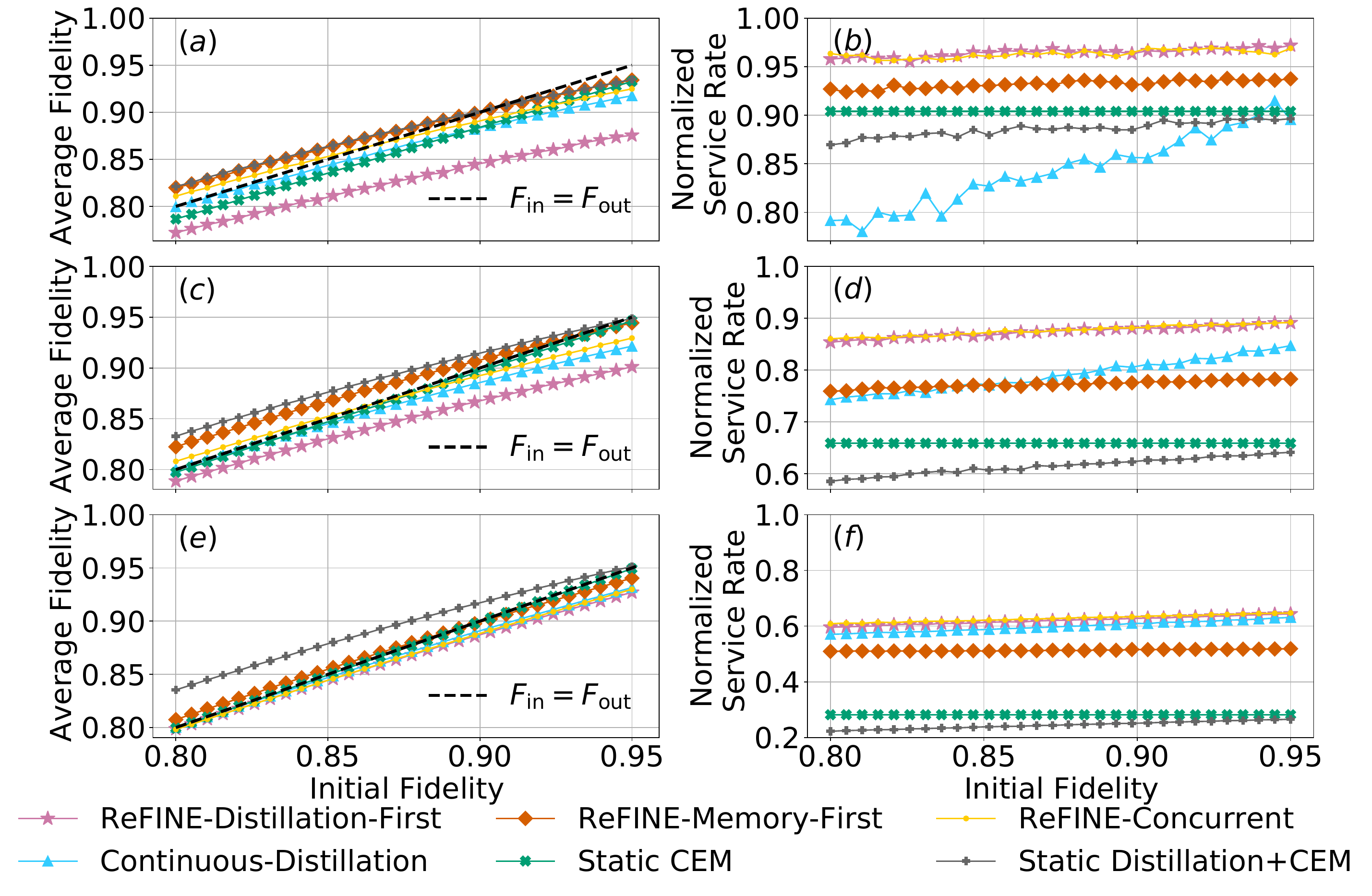}
    \caption{Sensitivity to initial EPR fidelity. (a–b) $\mu = 0.02\lambda$: \approachOne{} maintains highest fidelity and high rate. (c–d) $\mu = 0.1\lambda$: \methodName{} improves fidelity and preserves rate gap. (e–f) $\mu = 0.5\lambda$: fidelity gap narrows; rate gap remains consistent.}
    \label{fig:sensitivity_SF}
\end{figure}

Figures~\ref{fig:sensitivity_SF} demonstrate \methodName{}'s consistent advantage over static policies for varying initial EPR fidelity. In Figures~\ref{fig:sensitivity_SF}(a), \ref{fig:sensitivity_SF}(c), and \ref{fig:sensitivity_SF}(e), policies with points above the dashed line $F_{in} = F_{out}$ not only maintain, but also improve initial fidelity. Note that, without distillation or coding, decoherence significantly degrades fidelity (e.g., it drops from $F_{in} = 0.8$ to $0.41$ for $\mu = 0.02\lambda $). Therefore, overcoming decoherence just to maintain the initial fidelity is still a worthy improvement. ReFINE-D and ReFINE-C maintain higher service rates across all initial fidelity levels and consumption regimes, ideal for rate-priority applications. ReFINE-M consistently outperforms Static Distill+CEM in service rate with comparable fidelity, though higher initial fidelity narrows the fidelity gap.

\noindent{\bf{Resource efficiency: }} ReFINE-D uses CEM more sparingly than ReFINE-C while achieving similar performance, making it optimal for resource-constrained deployments. This demonstrates how demand-aware scheduling adapts both to consumption patterns and resource availability.

\subsubsection{$\Gamma_{CEM}$}\label{apx:gammaSen} 
\begin{figure}[h]
    \centering
    \includegraphics[width=0.95\columnwidth]{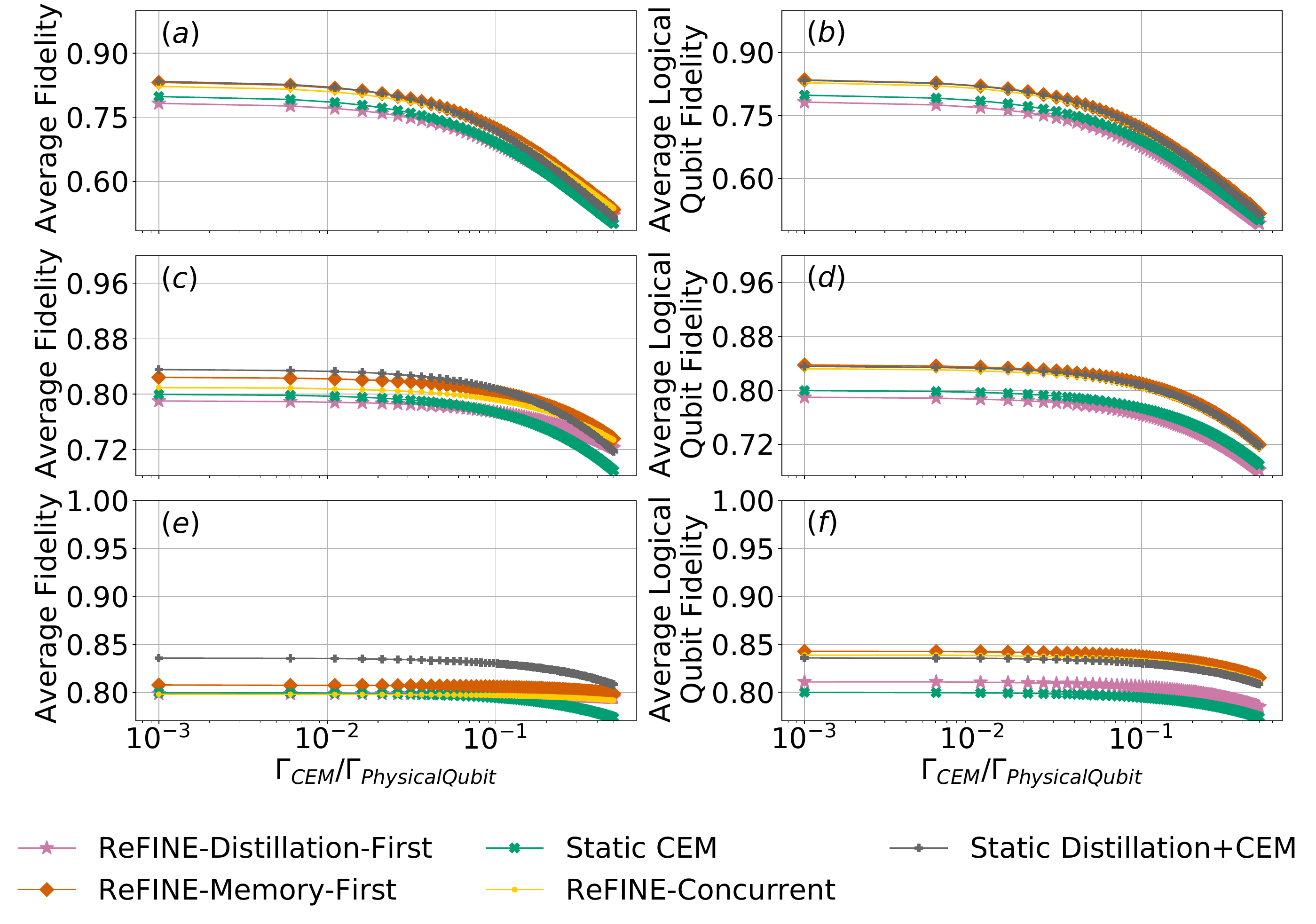}
    \caption{Sensitivity to CEM/physical-qubit $\Gamma$ ratio. Static policies degrade more at higher
ratios. (a,c,e)~Total average served fidelity; \methodName{}'s use of physical qubits
at high $\Gamma$ ratios makes it more robust. (b,d,f)~Fidelity of CEM-served EPR pairs
only; all policies are affected equally. Low request rates (a–b) are most sensitive to
CEM's $\Gamma$; high rates (e–f) are most robust as CEM is used less frequently.}
\label{fig:gamma_sensitivity}
\end{figure}

We analyze system performance across different CEM protection ratios ($\Gamma_{CEM}/\Gamma_{PhysicalQubit}$) to understand how CEM effectiveness impacts scheduling decisions. While our main evaluation assumes $\Gamma_{CEM} = 0.01 \times \Gamma_{PhysicalQubit}$, Figure~\ref{fig:gamma_sensitivity} shows results for ratios ranging from 0.001 to 0.1, with $\Gamma_{PhysicalQubit} = 0.05$.

As the $\Gamma$ ratio increases (indicating weaker CEM protection), all methods show fidelity degradation, especially for ratios higher than $0.1$. Static CEM and Static Distill+CEM show higher sensitivity since they rely exclusively on CEM for serving. ReFINE's policies demonstrate more robust performance across protection levels due to their hybrid serving approach. As consumption rates increase, ReFINE serves more requests from physical qubits due to frequent arrivals, naturally reducing exposure to CEM protection limitations. This adaptive behavior means ReFINE's performance degrades more gracefully as CEM protection weakens compared to Static CEM and Static Distill+CEM, which rely exclusively on CEM for serving, showing steeper performance drops with reduced protection, especially at higher consumption rates. 

\noindent{\bf{\new{Discussion on Different Codes and Distillation Protocols}}}
\new{Our evaluation instantiates CEM with the teleportation-based 9-qubit Shor code and distillation with the 2-to-1 Bilocal Clifford protocol, but ReFINE's scheduling rationale does not depend on either choice. ReFINE rests on two structural properties that hold for any reasonable code and distillation protocol: forming a code incurs a formation-time-versus-fidelity tradeoff (it takes time and resources to build the logical memory, during which requests may arrive), and distillation incurs a success-probability-versus-fidelity tradeoff (each round raises fidelity but may fail and destroy the pair). Crucially, a stronger code or distillation offers better protection but typically requires more entangled qubits to form, consuming more generator capacity and formation time. Therefore, a stronger code or protocol does not escape the rate-fidelity tradeoff.}
\subsection{Overhead of \methodName{}}\label{sec:overhead}
\new{Encoding EPR pairs into CEM incurs a cost, but \methodName{} schedules it, only when beneficial. Table~\ref{tab:cost} shows the cost of different policies for $\mu=0.1\lambda$. \methodName{} uses fewer average coding qubits used for serving, $N_{\textit{coding qubits}}$, than static methods. The normalized discarded EPR pair rate ($\lambda_{discarded}$), which occurs due to unsuccessful distillation, is significantly lower for \methodName{} (10–12\%) compared to Continuous Distillation (72\%). Additionally, the average number of consumed EPR pairs per service ($N_{\textit{Cons. EPR}}$) is lower for \methodName{} than prior distillation-based methods.
We assume decoding occurs in the background, similar to prior work~\cite{muralidharan2016optimal,jiang2009quantum,fowler2010surface,muralidharan2014ultrafast,munro2012quantum,patil2024entanglement, pathumsoot2024boosting, pattison2025constant} where the focus is not on decoding overhead.}
\vspace*{-5mm}
\begin{table}[h]
    \centering
    \footnotesize
    \caption{Cost of Coding and Distillation}
    \label{tab:cost}
    \renewcommand{\arraystretch}{1}
    \setlength{\tabcolsep}{6pt}
    \begin{threeparttable} 
    {
    \begin{tabular}{|l|P{1.5cm}|P{1.1cm}|P{1.4cm}|}
        \hline
        \textbf{Method} & \textbf{$N_{\textit{coding qubits}}$\tnote{1}} & \textbf{$\lambda_{discarded}$} & \textbf{$N_{\textit{Cons. EPR}}$} \\
        \hline \hline
        \textbf{\approachZero{}} & 6.2 & 0.12 & 2.6 \\
        \hline
        \textbf{\approachOne{}} & 7.5 & 0.10 & 2.6 \\
        \hline
        \textbf{\approachTwo{}} & 6.4 & 0.11 & 2.4 \\
        \hline
        \textbf{Static Distill+CEM.} & 10 & 0.10 & 3 \\
        \hline
        \textbf{Static CEM} & 10 & 0 & 1 \\
        \hline
        \textbf{Continuous-Distill.} & 0 & 0.72 & 3.3 \\
        \hline
    \end{tabular}
    }
    \begin{tablenotes}
        \footnotesize
        \item[1] \textit{9 (CEM) + 1 (Initialization through teleportation)}
    \end{tablenotes}
    \end{threeparttable}
\end{table}
\vspace*{-0.5cm}
\subsection{Ablation Study}\label{apx:ablation}

To isolate the impact of ReFINE's key scheduling components, we analyze how different scheduling decisions contribute to overall performance improvements.

\subsubsection{Impact of CEM Scheduling Policies}\label{sec:eval_CEM_Usage}

\begin{figure}[h]
    \centering
    \includegraphics[width=0.9\columnwidth]{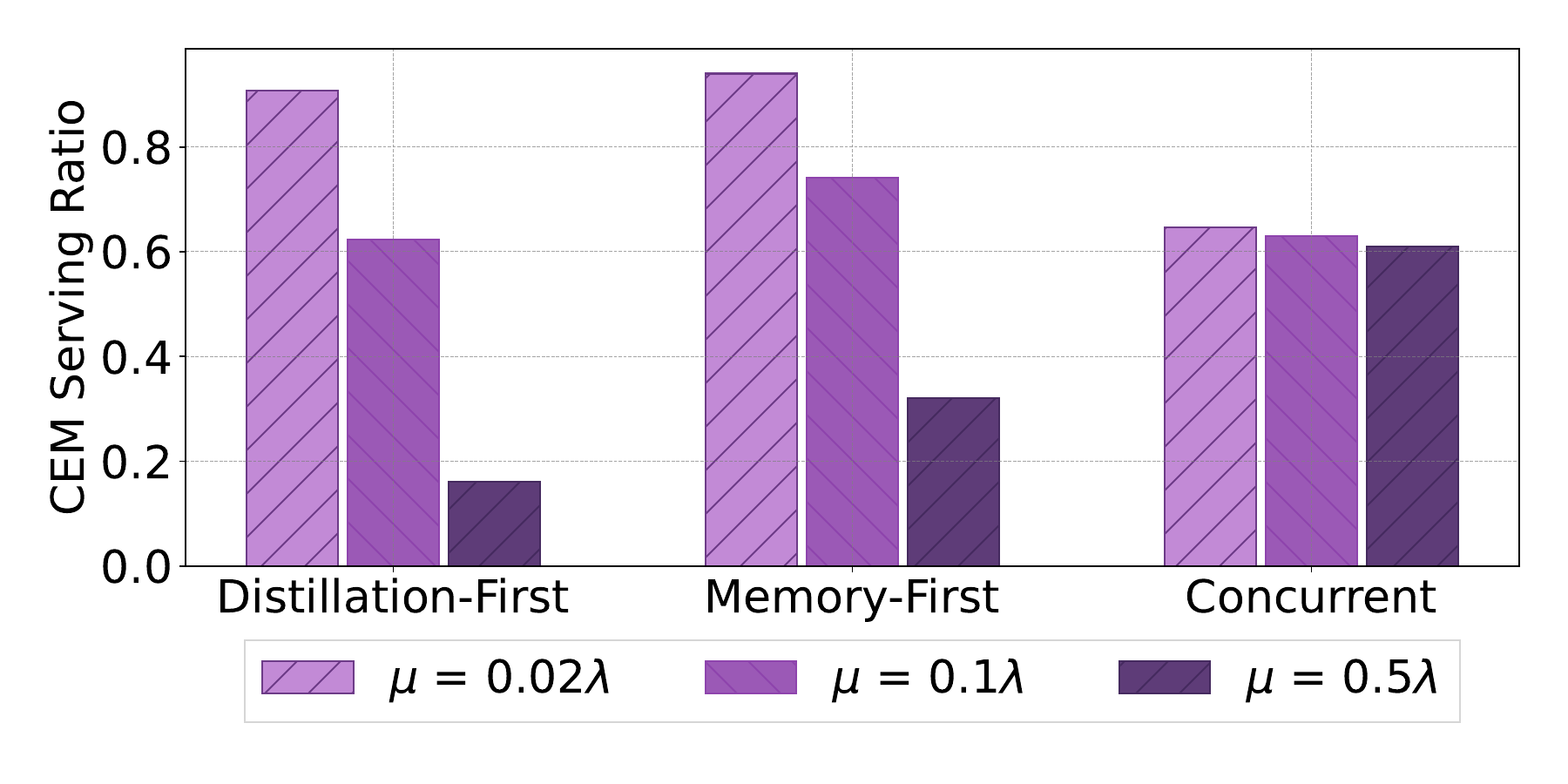}
    \caption{CEM usage ratio across scheduling policies under varying request rates.
    Higher request rates reduce the chance to complete CEM formation, leading to
    lower CEM usage.}
    \label{fig:cem_Ratio}
\end{figure}

Figure~\ref{fig:cem_Ratio} illustrates how request rate affects \methodName{}'s CEM usage,
highlighting the advantage of demand-aware scheduling. \approachOne{} achieves the highest
CEM usage at low-to-moderate rates ($\mu = 0.02\lambda$, $0.1\lambda$), as its early CEM
formation strategy maximizes preservation opportunities; however, it reduces CEM usage at
higher demand ($\mu = 0.5\lambda$), preemptively serving requests when CEM overhead becomes
counterproductive. \approachZero{} shows the most aggressive CEM reduction with increasing
$\mu$, prioritizing immediate serving, while \approachTwo{} maintains moderate, stable CEM
usage across all regimes. This dynamic adaptation is precisely what static policies cannot
achieve.

\subsubsection{Preemptive serving impact} Comparing \approachOne{} with
Static Distill+CEM (Section~\ref{sec:tradeoff}) shows the impact: both follow identical operation sequences but differ only in request
handling. Static Distill+CEM rejects all requests until distillation and CEM formation
fully complete, whereas \approachOne{} serves requests during intermediate stages.
\subsubsection{distillation vs.\ CEM} This is assessed by
comparing CED (distillation-only) against Static CEM (CEM-only), establishing the
performance range when schedulers are restricted to a single technique,  quantified in Section~\ref{sec:tradeoff}.
\subsubsection{Impact of Technique Orders in Policies}
Table~\ref{tab:CEM_Impact} analyzes how scheduling order affects fidelity at each  stage.

\noindent\textbf{Preserved $\Bar{F}$:} ReFINE-M and ReFINE-C preserve distilled pairs
immediately after distillation, since CEM formation completes first in both policies,
avoiding additional decoherence. ReFINE-D delays CEM formation until after distillation,
exposing EPR pairs to further decoherence before preservation.

\noindent\textbf{\LogicalQ{} Serving $\Bar{F}$:} Slightly lower than Preserved
$\Bar{F}$, reflecting CEM's own (much slower) decoherence rate.

\noindent\textbf{Physical Qubit Serving $\Bar{F}$:} \approachZero{} achieves the highest
value by serving distilled pairs directly from physical qubits before CEM formation,
whereas \approachOne{} and \approachTwo{} incur additional wait-induced decoherence.

These results confirm that ReFINE's gains stem from three scheduling innovations:
(1)~preemptive serving eliminates unnecessary rejections, (2)~policy selection targets
different application requirements, and (3)~timing decisions minimize decoherence during
resource transitions.
\vspace*{-0.5cm}
\begin{table}[h]
    \caption{Fidelity of Each Stage of \methodName{}}
    \centering
    \footnotesize
    \setlength{\tabcolsep}{0.2cm}
    \renewcommand{\arraystretch}{0.85}
    \begin{tabular}{|l|c|c|c|c|}
        \hline
         & $\mu$ & \makecell{\approachZero{}} & \makecell{\approachOne{}} & \makecell{\approachTwo{}} \\
        \hline
        \multirow{3}{*}{\makecell[tl]{Preserved $\Bar{F}$}} 
            & \centering $0.02\lambda$ & $0.783$ & $0.836$ & $0.830$ \\ 
            & \centering $0.1\lambda$  & $0.790$ & $0.838$ & $0.832$ \\ 
            & \centering $0.5\lambda$  & $0.811$ & $0.842$ & $0.838$ \\ 
        \hline
        \multirow{3}{*}{\makecell[tl]{\LogicalQ{}\\Serving $\Bar{F}$}} 
            & \centering $0.02\lambda$ & $0.770$ & $0.821$ & $0.816$ \\ 
            & \centering $0.1\lambda$  & $0.787$ & $0.835$ & $0.829$ \\ 
            & \centering $0.5\lambda$  & $0.810$ & $0.842$ & $0.838$ \\ 
        \hline
        \multirow{3}{*}{\makecell[tl]{Phys. Qubit\\Serving $\Bar{F}$}} 
            & \centering $0.02\lambda$ & $0.787$ & $0.783$ & $0.778$ \\ 
            & \centering $0.1\lambda$  & $0.790$ & $0.784$ & $0.775$ \\ 
            & \centering $0.5\lambda$  & $0.796$ & $0.790$ & $0.760$ \\ 
        \hline
    \end{tabular}
    \label{tab:CEM_Impact}
\end{table}

\vspace*{-0.45cm}
\subsection{\new{Swapping: Chain of Nodes}} 
\begin{figure}[h]
    \centering
    \includegraphics[width=\columnwidth]{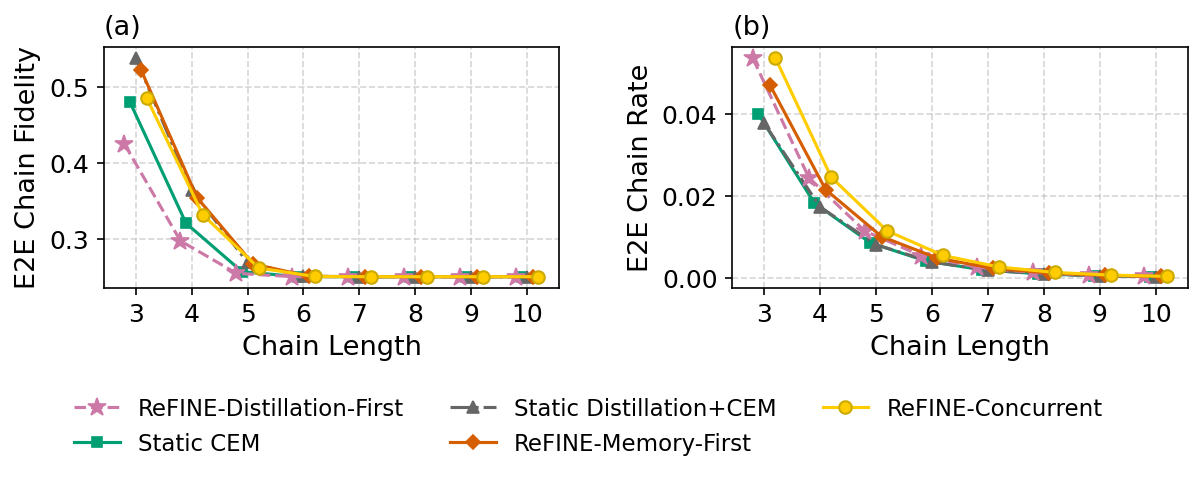}
    \caption{\new{End-to-end (E2E) chain fidelity (a) and EPR distribution rate (b) as a function of chain length. ReFINE outperforms prior works in short chains.}}
    \label{fig:chain}
\end{figure}

\new{ReFINE improves the rate–fidelity tradeoff over the static baselines at short chains, where per-link fidelity gains carry through to the end-to-end pairs. As the chain grows, illustrated in Figure~\ref{fig:chain}(a-b), entanglement-swapping errors accumulate across hops and drive the end-to-end fidelity of all methods toward the maximally-mixed floor, so the policies converge to a similarly inefficient point. Per-link scheduling such as ReFINE therefore helps most on short-to-moderate chains, while long chains are swap-error-limited.}

%% file: sections/6_related_work.tex
\section{Related Work}

\noindent \textbf{Network-Wide Entanglement Distribution:} To extend the range of quantum communication, quantum repeaters are employed as essential devices. They create entangled links between shorter segments and perform processes such as entanglement swapping and distillation to maintain high fidelity~\cite{briegel1998quantum,muralidharan2016optimal}. This network-wide distribution of entangled states also requires efficient routing. Prior work~\cite{PhysRevA.100.032322,chehimi2024entanglement,pouryousef2024resource,inesta2023optimal,li2021efficient, davies2024tools, praxmeyer2013reposition,patil2024entanglement,pathumsoot2024boosting,zhang2025switchqnet} works on establishing end-to-end entangled links across a series of quantum repeaters and switches for wide ranges of applications. In contrast to these studies, \methodName{} focuses on link-level entanglement distribution, a more immediate challenge that paves the way for applications such as QKD.

\noindent\textbf{Scheduling in Quantum Networks:} Prior work has addressed scheduling challenges at multiple levels, including network-wide quantum repeater placement~\cite{pouryousef2024resource, gyongyosi2020resource,van2008system}, resource allocation for switches~\cite{vasantam2022throughput,cicconetti2021request}, end-to-end request scheduling through path-finding and swapping~\cite{vasantam2021stability,wang2023efficient,sundaram2024optimized, li2021effective}, and network-wide distillation scheduling~\cite{xiao2024purification}. These approaches focus on which requests to serve or which network paths to use, but do not address how individual nodes should sequence their local operations. Specifically, no prior work addresses the temporal dependencies between distillation and CEM formation within nodes, the timing-dependent utility of error correction, or preemptive serving during intermediate processing stages. ReFINE fills this gap by providing the first node-level scheduler that optimizes operation ordering and resource allocation timing based on demand patterns, complementing existing scheduling approaches.

\noindent\textbf{QEC in Entanglement Distribution:}
QEC in entanglement distribution can correct operational errors~\cite{pattison2025constant,jiang2009quantum,munro2010quantum,fowler2010surface,muralidharan2014ultrafast} using only one-way classical signaling. This approach enables distributed FTQC but represents a long-term vision. We focus on more near-term quantum networks where QEC mitigates decoherence effects. Scheduling for distributed FTQC remains as future work.

%% file: sections/8_conclusion.tex
\section{Discussion and Conclusion}

This work introduces \methodName{}, a scheduler that enables significantly better balance for rate-fidelity tradeoff in quantum networks by adapting to network demands. Unlike static methods that commit to fixed sequential processing regardless of request timing, ReFINE dynamically schedules distillation and CEM formation while serving requests preemptively based on consumption request arrivals.

Through three policies, Distillation-first, Memory-first, and Concurrent, \methodName{} allocates resources between distillation and CEM formation, according to varying fidelity and rate requirements of the network. Note that  the distillation protocol used here has been widely studied and, to our knowledge, the only one experimentally implemented~\cite{kalb2017entanglement,yan2022entanglement}. However, the protocol has a theoretical upper bound beyond which fidelity improvements diminish; for an initial fidelity of 0.8, the maximum achievable fidelity is only 0.87. Evaluations show that \methodName{} closes 32\% of this gap, reaching a fidelity of 0.823. Compared to ideal fidelity 1 (theoretically unattainable), \methodName{} closes up to 11.21\% of the gap over Continuous Distillation and up to  49.48\% of the gap to ideal service rate, where request serving matches arrivals, over Static Distill+CEM. Thus, the improvement is not minor relative to the room for enhancement. Therefore, \methodName{} paves the way for more resource-efficient and performance-optimized quantum networks, addressing the critical tradeoffs inherent in current architectures.

%% file: sections/acknowledgement.tex
\section*{Acknowledgement}
We thank anonymous reviewers of HPCA-2025, ISCA2025, MICRO-2025, and ASPLOS-2026 for their feedback.

%% file: sections/9_appendix.tex
\appendix
\subsection{Fidelity and Probability of Success for 2-to-1 Entanglement Distillation}\label{apx:distillation}

We use 2-to-1 protocol~\cite{deutsch1996quantum} for distillation. This distillation protocol used here has been widely studied in quantum network, and to our knowledge, the only one experimentally implemented~\cite{kalb2017entanglement,yan2022entanglement}. For the distillation process two Bell-diagonal mixed states are used: one target pair (kept upon successful measurement), and one auxiliary pair (measured and discarded). Each state is assumed to be a Werner state with fidelity toward the Bell state $|\phi^+\rangle$.

Let the diagonal entries of the Bell-basis density matrix for the auxiliary and target pairs be given by:
\begin{align*}
\rho &= A|\phi^+\rangle\langle\phi^+| + B|\psi^-\rangle\langle\psi^-| + C|\psi^+\rangle\langle\psi^+| + D|\phi^-\rangle\langle\phi^-| \\
\rho' &= A'|\phi^+\rangle\langle\phi^+| + B'|\psi^-\rangle\langle\psi^-| + C'|\psi^+\rangle\langle\psi^+| + D'|\phi^-\rangle\langle\phi^-|
\end{align*}

Following the distillation procedure in Deutsch et al.~\cite{deutsch1996quantum}, when Alice and Bob retain the target pair (i.e., when their measurements on the auxiliary pair coincide), the resulting density matrix $\tilde{\rho}$ has diagonal elements:
\begin{align*}
\tilde{A} &= \frac{A A' + B B'}{N}, &
\tilde{B} &= \frac{C D' + C' D}{N}, \\
\tilde{C} &= \frac{C C' + D D'}{N}, &
\tilde{D} &= \frac{A B' + A' B}{N},
\end{align*}
where the normalization factor $N$ is the probability of success (i.e., matching measurement outcomes):
\[
N = (A + B)(A' + B') + (C + D)(C' + D').
\]

\subsection*{Werner State Substitution} Assuming Werner states for target and auxiliary pairs are
\begin{align*}
\small\rho_{w_{target}} = &\;F\ket{\phi^+}\bra{\phi^+} + \\
&\;\frac{1-F}{3} ( \ket{\psi^+}\bra{\psi^+} + \ket{\psi^-}\bra{\psi^-}
        + \ket{\phi^-}\bra{\phi^-} )
\end{align*}

\begin{align*}
      \small\rho_{w_{aux}} = &\;F_{aux}\ket{\phi^+}\bra{\phi^+} +\\ &\;\frac{1-F_{aux}}{3} ( \ket{\psi^+}\bra{\psi^+} + \ket{\psi^-}\bra{\psi^-}
        + \ket{\phi^-}\bra{\phi^-} ).
\end{align*}

So, we substitute:
\begin{align*}
A &= F, &
B = C = D &= \frac{1 - F}{3}, \\
A' &= F_{\text{aux}}, &
B' = C' = D' &= \frac{1 - F_{\text{aux}}}{3}.
\end{align*}

We compute:
\begin{align*}
A + B &= F + \frac{1 - F}{3} = \frac{2F + 1}{3}, \\
C + D &= 2 \cdot \frac{1 - F}{3} = \frac{2(1 - F)}{3}, \\
A' + B' &= \frac{2F_{\text{aux}} + 1}{3}, \\
C' + D' &= \frac{2(1 - F_{\text{aux}})}{3}.
\end{align*}

Thus, the success probability becomes:
\begin{align*}
P_{\text{succ}} &= N = (A + B)(A' + B') + (C + D)(C' + D') \\
&= \left( \frac{2F + 1}{3} \right) \left( \frac{2F_{\text{aux}} + 1}{3} \right)\\
 &+ \left( \frac{2(1 - F)}{3} \right)\left( \frac{2(1 - F_{\text{aux}})}{3} \right) \\
&= \frac{(2F + 1)(2F_{\text{aux}} + 1) + 4(1 - F)(1 - F_{\text{aux}})}{9}.
\end{align*}

\subsection*{Fidelity After One Round}

The updated fidelity $\tilde{F} = \tilde{A}$ is given by:
\begin{align*}
\tilde{F} &= \frac{F F_{\text{aux}} + \left( \frac{1 - F}{3} \right) \left( \frac{1 - F_{\text{aux}}}{3} \right)}{P_{\text{succ}}} \\
&= \frac{9F F_{\text{aux}} +(1 - F)(1 - F_{\text{aux}})}{9P_{\text{succ}}}.
\end{align*}
This function characterizes the output fidelity after one round of successful distillation, given the initial fidelities of the target and auxiliary pairs.

\vspace{0.5cm}
\subsection{Markov Chain Models for \methodName{}'s Policies}\label{apx:design-markov}
\begin{figure}[t!]
    \centering
    \includegraphics[width=\columnwidth]{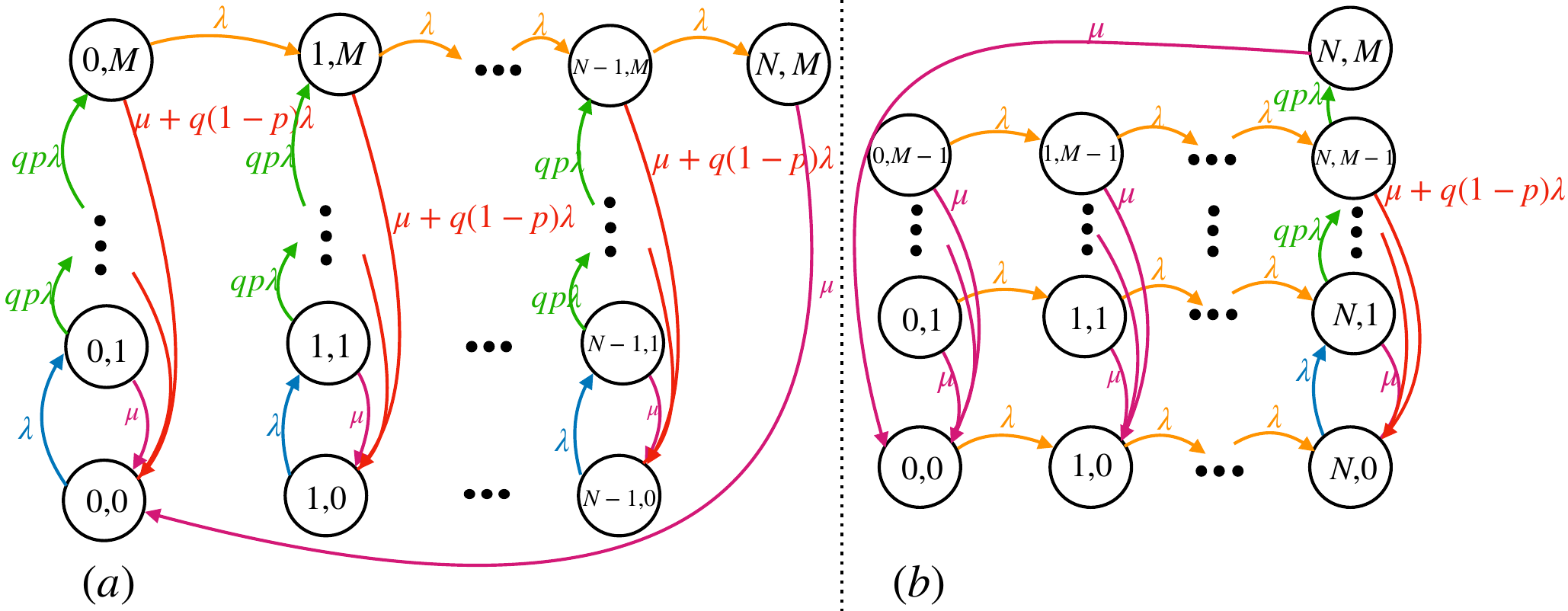}
    \caption{(a) \approachZero{}'s Markov chain model.  \approachZero{} starts with dedicating EPR pair generator to distillation.
    (b)~\approachOne{}'s Markov chain model.  \approachOne{} prioritizes employing EPR pair generator for \LogicalQ{} formation. }
    \label{fig:markov_refineDandM}
\end{figure}

\begin{figure}[h]
    \centering
    \includegraphics[width=0.7\columnwidth]{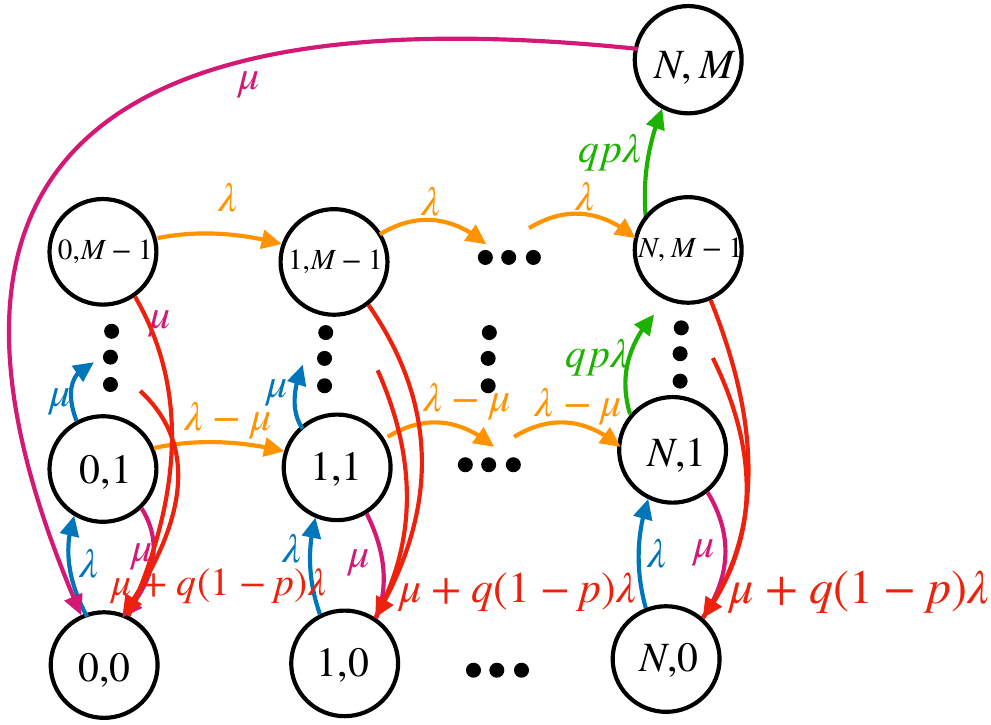}
    \caption{Markov chain model for \approachTwo{}. After the first distribution, it concurrently  distills, with rate $\mu$, and forms \LogicalQ{}, with rate $\lambda - \mu$. After completing the CEM formation, whole generation rate $\lambda$ is utilized for distillation.}
    \label{fig:adaptive}
\end{figure}

We define \methodName{}'s system and policy for entanglement generation and distillation as a Markov chain. We consider the state of the system at any time \(t\) as \(s(t) = (i,j)\), where \(i\) represents the number of EPR pairs Sender has prepared for \LogicalQ{}. Here, \(i \in \{0, 1, ..., N\}\) and \(N\) is the number of EPR pairs needed for \LogicalQ{}. In our framework, \LogicalQ{} is based on Shor's code~\cite{luo2021quantum} of which \(N = 2\). The index $j$ in $(i, j)$ tracks the target qubit and indicates whether the target EPR pair is not shared ($j=0$) or if it is shared with Receiver ($j=1$). An index $1 < j \leq M$ means that the target EPR pair is distilled $j-1$ rounds. The distillation success probability of each round is~$p$. Scheduler uses a generated EPR pair with probability $q$ to be consumed for distillation, $q$ is a hyperparameter to tune the fidelity and service rate. The EPR pair generator creates an EPR pair according to a Poisson process with rate $\lambda$. EPR pair consumption requests are based on a Poisson process with rate $\mu$. Table~\ref{tab:notation} summarizes the parameters used in our work. Figure~\ref{fig:markov_refineDandM}(a),~\ref{fig:markov_refineDandM}(b), and~\ref{fig:adaptive} show the Markov chain models for \approachZero{}, \approachOne{}, and \approachTwo{}, respectively. \new{EPR pairs allocated for CEM formation may be lost with probability $ 0.01$
prior to CEM initialization}, omitted from the
figure to avoid clutter.
\vspace{0.4cm}
\begin{table}[h!]
    \caption{Parameters of \methodName{}}
    \centering
    \footnotesize
    \setlength{\tabcolsep}{0.1cm} 
    \renewcommand{\arraystretch}{1.2} 
    \small 
    \begin{tabular}{| c | p{6.8cm} |} 
        \hline
        Parameter & Definition \\
        \hline \hline
        $\lambda$ & EPR pair generation rate (Sender's side) \\ \hline
        $\mu$ & Consumption request rate \\ \hline
        $p$ & Distillation success probability
        \\ \hline
        $q$ & Hyperparameter for tuning fidelity and service rate\\ 
        \hline
        $(i,j)$ & ($\#$ of EPR pairs required for \LogicalQ{} formation, $\#$ of Distillation rounds - 1).  \\ 
        \hline
        $M$ & Maximum number of distillation rounds - 1 \\\hline
        $N$ & Number of EPR pairs required for coding \\\hline
    \end{tabular}
    \label{tab:notation}
\end{table}

\vspace{0.4cm}
\newpage